\documentclass[showpacs,showkeys,preprint]{revtex4}

\usepackage[latin1]{inputenc}
\usepackage[T1]{fontenc}
\usepackage{amsmath,amssymb}
\usepackage{array}
\usepackage{subfigure}
\usepackage[draft]{graphicx}

\begin{document}

\newcommand{\re}{\text{Re}}
\newcommand{\im}{\text{Im}}
\newcommand{\sign}{\text{sign}}
\newcommand{\de}{\mbox{d}}
\newcommand{\eref}[1]{(\ref{#1})}
\newcommand{\relphantom}[1]{\mathrel{\phantom{#1}}}
\newcommand{\ii}{\mbox{i}}
\def\reff#1{(\ref{#1})}

\newcommand {\be}{\begin{equation}}
\newcommand {\ee}[1]{\label{#1}\end{equation}}

\newcommand {\rem}[1]{}

\title{Compactons and Chaos in Strongly Nonlinear Lattices}
\date{\today}

\author{Karsten Ahnert}
\author{Arkady Pikovsky}
\affiliation{Department of Physics and Astronomy, Potsdam University, 14476 Potsdam, Germany}

\begin{abstract}
We study localized traveling waves and chaotic states in strongly nonlinear
one-dimensional Hamiltonian lattices. We show that the solitary waves are
super-exponentially localized, and present an accurate numerical method
allowing to find them for an arbitrary nonlinearity index. Compactons evolve
from rather general initially localized perturbations and collide nearly
elastically, nevertheless on a long time scale for finite lattices an
extensive chaotic state is generally observed. Because of the system's
scaling, these dynamical properties are valid for any energy.
\end{abstract}

\pacs{05.45.Yv, 63.20.Ry, 05.45.-a}
\keywords{Nonlinear Hamiltonian lattices; Compactons; Space-time chaos}

\maketitle

\section{Introduction}
\label{sec:intro}
Hamiltonian lattices are one of the simplest objects in nonlinear physics,
nevertheless they still elude full understanding. Already the first attempt to
understand nonlinear effects ended with the Fermi-Pasta-Ulam puzzle, which is
still not fully resolved~(see e.g. the focus issue on ``The
``Fermi-Pasta-Ulam" problem -- the first 50 years'' in~\cite{Chaos-fpu-05});
another remarkable feature found only recently is the existence of localized
breathers~\cite{Flach-Willis-98}. Quite often nonlinear effects in lattices
can be treated perturbatively, leading to well-established concepts of phonon
interaction and weak turbulence. Beyond a perturbative account of a weak
nonlinearity one is encountered with genuine nonlinear phenomena, like
solitons and chaos. The level of nonlinearity usually grows with the energy,
allowing one to follow a transition from linear to nonlinear regimes by
pumping more energy in the lattice.

In this paper we study strongly nonlinear Hamiltonian lattices that do not
possess linear terms. We restrict our attention to the simplest
one-dimensional case where particles interact nonlinearly and no on-site
potential is present.  We choose the interaction potential in the simplest
power form, thus the lattice is characterized by a single parameter -- the
nonlinearity index. The equations of motion obey scaling, what means that the
dynamical properties are the same for all energies -- only the time scale
changes. Lattices of this type attracted large attention recently, in
particular due to a prominent example -- the Hertz lattice that describes
elastically interacting hard balls, it has the nonlinearity index
$3/2$~\cite{Nesterenko-83,Coste-Falcon-Fauve-97,Chatterjee-99,Porter-08}. We
focus our study on the interplay of solitary waves and chaos in such
lattices. Some 25 years ago
Nesterenko~\cite{Nesterenko-83,Lazaridi-Nesterenko-85,Gavrilyuk-Nesterenko-94,Nesterenko-01}
has described a compact traveling wave solution in the Hertz lattice, which
can be understood as a compacton. Compactons have been introduced, in a
mathematically rigor form, by Rosenau and
coworkers~\cite{Rosenau-Hyman-93,Rosenau-94} for a class of nonlinear PDEs
with nonlinear dispersion. Compactons can be analytically found if one
approximates the lattice equations with nonlinear PDEs, but less is known on
the genuine lattice solutions. Below, in Section~\ref{sec:twla} we present a
numerical procedure for determining traveling waves for an arbitrary
nonlinearity index, and compare these solutions with those of the approximated
PDEs (Section~\ref{subsec:qca}). Furthermore, we show that compactons
naturally appear from localized initial perturbations and relatively robustly
survive collisions, but nevertheless evolve to chaos on a long time scale in
finite lattices (Section~\ref{sec:evcol}). Properties of chaos are studied in
Section~\ref{sec:ch}. We demonstrate extensivity of the chaotic state by
calculating the Lyapunov spectrum, and study the dependence of the Lyapunov
exponents on the nonlinearity index. Some open questions are discussed in the
concluding Section~\ref{sec:con}.

\section{The model}
\label{sec:model}

Our basic model is a family of lattice Hamiltonian systems
\begin{equation}
H = \sum_k \frac{p_k^2}{2} + \frac{1}{n+1}|q_{k+1} - q_k|^{n+1}
\label{eq:ham} \textrm{,}
\end{equation}
which are parameterized by one real parameter -- the nonlinearity index $n$.
Below we assume that $n\geq 1$.  The case $n=1$ corresponds to a linear
lattice. Another interesting case is $n=3/2$. Such a nonlinearity appears,
according to the Hertz law, at the compression in a chain of elastic hard
balls. For the realistic system of balls, however, the potential has the form
like in \reff{eq:ham} only for $q_{k+1} - q_k<0$, for $q_{k+1} - q_k>0$ no
attracting force is acting. A simplified realization of such a system is the
toy ``Newton's cradle'', which possesses the same Hertzian interaction
law. However the standard ``Newton's cradle'' consists of a few balls
(typically 5), which are not enough for the formation of stationary traveling
waves. Furthermore, slight intervals between adjacent beads are not excluded,
contrary to experiments \cite{Lazaridi-Nesterenko-85,Coste-Falcon-Fauve-97}
where great care is taken to let the beads in effective contact. For different
aspects of the Hertz chain see
refs. \cite{Sinkovits-Sen-95,Nesterenko-01,Sen-Sinkovits-96,Sen-Manciu-Wright-98,Manciu00,Manciu-Sen-Hurd-00,Sen-Manciu-01,Hascoet-Hinch-02,Rosas-Lindenberg-03,Rosas-Lindenberg-04,Job-Melo-Sokolow-Sen-05,Porter-08},
a review article \cite{Sen-Hong-Bang-Avalos-Doney-08}, and references
therein. Contrary to this, in our model \reff{eq:ham} we assume both repulsive
and attracting forces.

Note that the potential in \reff{eq:ham} is generally non-smooth, except for
cases $n=1,3,5,\ldots$. Although the dynamics can be easily studied in
non-smooth situations as well, we will mainly focus below on the simplest
smooth non-trivial case $n=3$.

The lattice equation of motion read
\begin{equation}
\ddot{q_k} = \left|q_{k+1}-q_k\right|^n\sign(q_{k+1}-q_k) - \left|q_k
-q_{k-1}\right|^n\sign(q_k -q_{k-1}) \textrm{.} \label{eq:motion}
\end{equation}
Since on the right hand side of \eref{eq:motion} only differences enter, it is
convenient to introduce the difference coordinates $Q_k=q_{k+1}-q_k$. Then the
equations of motion are transformed to
\begin{equation}
\ddot{Q}_k= \left|Q_{k+1}\right|^n\sign(Q_{k+1}) -
2 \left|Q_k\right|^n\sign(Q_k) + 
\left|Q_{k-1}\right|^n\sign(Q_{k-1})
\label{eq:motion_diff} \textrm{.}
\end{equation}
Note, that a solitary wave in the variables $Q_k$ corresponds to a kink
(shock-like wave) in the variables $q_k$.

\paragraph{Conservation laws.}
The equations of motion possess two conservation laws: the energy and the
total momentum, the latter can be trivially set to zero by transforming into a
moving reference frame.

\paragraph{Scaling.} As mentioned in Refs.~\cite{Nesterenko-83,Nesterenko-01,Chatterjee-99}, the lattice \reff{eq:ham} has remarkable
scaling properties, due to homogeneity of the interaction energy. It is easy
to check, that the Hamiltonian can be rescaled according to \be q=a\tilde
q,\quad p=a^{\frac{n+1}{2}}\tilde p,\quad H=a^{n+1}\tilde H,\quad
t=a^{\frac{1-n}{2}}\tilde t.  \ee{eq:resc} Note, that this scaling involves
only the amplitude and the characteristic time of the solutions: by decreasing
the amplitude one obtains new solutions having the same spatial structure but
evolving slower. We will see that this property has direct consequences for
the properties of traveling waves and of chaos.

\section{Traveling solitary waves}

In this section localized traveling waves are investigated, first in a
quasicontinuous approximation (QCA) and then via numerical solution of the
lattice equations.  A mathematically rigor proof of the existence of
  solitary waves in Hamiltonian lattices of type \eref{eq:ham} has been given in 
Refs.~\cite{Friesecke-Wattis-94,MacKay-99}.

\subsection{Quasicontinuous approximation}
\label{subsec:qca}

Here, we represent the solution of the lattice equations \reff{eq:motion_diff}
as a function of two continuous variables $Q(x,t)$. We are seeking for
solitary waves which do not change their sign, for definiteness we consider
$Q\geq 0$ (this consideration is therefore suitable for lattices where the
nonlinearity index is different for positive and negative displacements $Q$,
e.g. for the Hertz lattice of elastic balls).  We present two approaches to
find a continuous version of the lattice. In the first one, we approximate the
differences between two displacements $Q$, while in the second one the
displacement $q$ at each lattice site is expanded directly.

\subsubsection{Expansion of differences}
Here we look for a direct quasicontinuous approximation of Eq.~\reff{eq:motion_diff}.
Expanding the difference coordinates $Q_k$ up to the fourth order we obtain
\begin{equation}
Q_{k \pm 1}^n \approx  Q^n(x,t) \pm h \left[Q^n(x,t)\right]_x + 
\frac{h^2}{2} \left[Q^n(x,t)\right]_{xx}
\pm \frac{h^3}{6} \left[Q^n(x,t)\right]_{xxx}
+ \frac{h^4}{24} \left[Q^n(x,t)\right]_{xxxx}
\textrm{,} \label{eq:qca_base_approx}
\end{equation}
where $h$ is the spatial difference between two lattice sites and
the subscripts denote differentiation with respect to $x$. Inserting
\eref{eq:qca_base_approx} into \eref{eq:motion_diff} and setting $h=1$ one
arrives at the partial differential equation
\begin{equation}
[Q(x,t)]_{tt} = \left[Q^n(x,t)\right]_{xx} + \frac{1}{12}
\left[Q^n(x,t)\right]_{xxxx} \textrm{.} \label{eq:qca}
\end{equation}
Equation~\reff{eq:qca} belongs to a class of strongly nonlinear PDEs, because
the dispersion term with the fourth derivative is nonlinear. The equation does
not possess linear wave solutions (this situation has been called ``sonic
vacuum'' by V. Nesterenko), 
but it has nontrivial nonlinear ones. In this way it is very
similar to a family of strongly nonlinear generalizations of the
Korteveg-de Vries equation, studied in~\cite{Rosenau-Hyman-93} and can be
considered as a strongly nonlinear version of the Boussinesq
equation~\cite{Rosenau-94}.

Now we seek for traveling wave solutions of \eref{eq:qca}, by virtue of the
ansatz \be Q(x,t)=Q(x-\lambda t)=Q(s).  \ee{eq:twa} Then~\reff{eq:twa} reduces
to the ODE
\begin{equation}
\lambda^2  Q_{ss} = \left[Q^n\right]_{ss} + \frac{1}{12}
\left[ Q^n \right]_{ssss} \textrm{.} \label{eq:qca_twa}
\end{equation}
Furthermore, we assume that the solution tends to zero as $s\to\pm\infty$, thus
after integrating twice we obtain
\begin{equation}
\lambda^2 Q = Q^n + \frac{1}{12} \left[ Q^n \right]_{ss} \textrm{.}
\label{eq:qca_twa_general}
\end{equation}
This equation also appears in the traveling waves ansatz for the
$K(n,n)$-equation in \cite{Rosenau-Hyman-93}.
Eq.~\eref{eq:qca_twa_general} can be solved for an arbitrary power $n$ by
\begin{subequations}
\label{eq:qca_twa_general_solution}
\begin{equation}
Q(s) = |\lambda|^m A_1 \cos^m (B_1s) 
\end{equation}
with
\begin{equation}
m = \frac{2}{n-1} \textrm{,} \qquad
A_1 = \left(\frac{n+1}{2n} \right)^{\frac{1}{1-n}} \textrm{,} \qquad
B_1 = \sqrt{3}\; \frac{n-1}{n} \textrm{.}
\end{equation}
\end{subequations}

\subsubsection{Expansion of displacements}

Another type of quasi-continuum can be obtained if we approximate
Eq.~\reff{eq:motion}. Now the displacement $q$ at each lattice site is written
as a continuous variable, what for the same order of the spatial derivative as
in \reff{eq:qca_base_approx} gives:
\begin{equation}
q_{k\pm1} = q \pm h q_x + \frac{h^2}{2} q_{xx} \pm \frac{h^3}{6} q_{xxx} +
\frac{h^4}{24} q_{xxxx} 
\label{eq:qca_base_approx_pre}
\end{equation}
Inserting this expansion into the equations of motion \eref{eq:motion} and
collecting all terms up to order of $h^{n+3}$ yields
\begin{equation}
[q]_{tt} = h^{n+1} [q_x^n]_x + \frac{h^{n+3}}{12}
\left( [q_x^n]_{xxx} - \frac{n(n-1)}{2} [ q_x^{n-2} q_{xx}^2]_x
\right)
\label{eq:qca_direct_pre} \text{.}
\end{equation}
This equation is the long wave approximation of Nesterenko
\cite{Nesterenko-83,Nesterenko-01}. To compare it with Eq.~\eref{eq:qca}, we
differentiate \eref{eq:qca_direct_pre} with respect to $x$, define $\tilde{Q}
= h q_x$ and set $h=1$:
\begin{equation} 
[\tilde{Q}]_{tt} = [\tilde{Q}^n]_{xx} + \frac{1}{12} \Big(
[\tilde{Q}^n]_{xxxx} - \frac{n(n-1)}{2} [\tilde{Q}^{n-2} \tilde{Q}_x^2]_{xx}
\Big)
\text{.} \label{eq:qca_direct}
\end{equation}
One can see that there is an additional term in \reff{eq:qca_direct} compared
to \eref{eq:qca}. This is not so much surprising, as these two
quasi-continuous approximations correspond to expansions at the different
positions of the original lattice, this effect is well-known for
approximations of Hamiltonian lattices with PDEs~\cite{Rosenau-03}. Because in
the problem we do not have a small parameter (the lattice spacing $h=1$ is not
small compared to the wave length), none of the equations
\reff{eq:qca},\reff{eq:qca_direct} can be expected to be exact in some
asymptotic sense. Instead, one has to justify them by comparing the solutions
with those of the full lattice problem, see Section~\ref{sec:twla} below.

To find traveling waves in the direct expansion we use again the ansatz
\eref{eq:twa} $\tilde Q(x,t) = \tilde Q(x-\lambda t) = \tilde Q(s)$. Inserting this ansatz and
integrating twice yields then an analogon to \eref{eq:qca_twa_general}
\begin{equation}
\lambda^2 \tilde Q = \tilde Q^n + \frac{1}{12} [\tilde Q^n]_{ss} - \frac{n(n-1)}{24} \tilde Q^{n-2}\tilde Q_s^2
\text{.} \label{eq:qca_twa_direct}
\end{equation}
One partial solution of this ordinary differential equation can also be
written as
\begin{subequations}
\label{eq:qca_twa_direct_solution}
\begin{equation}
\tilde Q(s) = |\lambda|^m A_2 \cos^m ( B_2 s )
\end{equation}
but with different constants $A_2$ and $B_2$
(cf.~\cite{Rosas-Lindenberg-03,Nesterenko-94})
\begin{equation}
m=\frac{2}{n-1}
\text{,} \qquad
A_2 = \left( \frac{2}{1+n} \right)^{\frac{1}{1-n}} 
\text{,} \qquad
B_2 = \sqrt{6\frac{(n-1)^2}{n(n+1)}}
\text{.}
\end{equation}
\end{subequations}

The solutions \reff{eq:qca_twa_general_solution} and
\reff{eq:qca_twa_direct_solution} do not satisfy boundary conditions,
moreover, they intersect with another, trivial solution of~\reff{eq:qca_twa}
$Q=0$. Remarkably, because of the degeneracy of Eq.~\reff{eq:qca_twa} and
Eq.~\reff{eq:qca_twa_direct} at zero, one can merge periodic solutions
\eref{eq:qca_twa_general_solution} and \eref{eq:qca_twa_direct_solution} with
the trivial solution $Q=0$ (see a detailed discussion
in~\cite{Rosenau-Hyman-93,Rosenau-94}):
\begin{equation}
Q(s) =
\begin{cases}
 |\lambda|^m A_i \cos^m (B_i s) & 
|s|<\frac{\pi}{2B_i} \\
0 & \textrm{else} \textrm{.}
\end{cases}
\label{eq:qca_twa_solution} 
\end{equation}
with $i=1,2$. This gives a compacton -- a solitary wave with a compact support
-- according to definition~\cite{Rosenau-Hyman-93,Rosenau-94}. For other,
non-solitary solutions of \eref{eq:qca_direct}, see
e.g.~\cite{Nesterenko-01,Nesterenko-94}. Note that due to the symmetries $x\to
-x$, $Q\to -Q$, solitary waves with both signs of velocity $\lambda$ and of
amplitude $A$ are the solutions.

It is important to check the validity of solution \reff{eq:qca_twa_solution}
by substituting it back to \reff{eq:qca_twa} or \reff{eq:qca_direct}. Then no
terms are singular for the case $m>2$ only, i.e. for $n<2$. Thus, the
constructed compacton solution \reff{eq:qca_twa_solution} is, strictly
speaking, not valid for strong nonlinearities $n\geq 2$.  This conclusion is,
however, only of small relevance for the original lattice problem. Indeed, the
PDE \reff{eq:qca} or \eref{eq:qca_direct_pre} is only an approximation of the
lattice problem: because the spatial extent of solution
\reff{eq:qca_twa_solution} is finite, there is no small parameter allowing us
to break expansion \reff{eq:qca_base_approx} or \eref{eq:qca_base_approx_pre}
somewhere. Just breaking it after the fourth derivative is arbitrary and can
be justified only by the fact, that in this approximation one indeed finds
reasonable solutions at least for some values of $n$. A real justification can
come only from a comparison with the solutions of the lattice equations, to be
discussed in the next subsection. And there we will see that the solution can
be found both for weak and strong nonlinearities $n>2$.

\rem{ We would like to mention that Eq.~\reff{eq:qca_base_approx} is not the
  only quasicontinuous approximation possible. Indeed, instead of performing
  the expansion of the variable $Q_k$, one could expand $q_k$ and insert this
  in \reff{eq:motion}, this would give a slightly different version of a PDE
  (see \cite{Rosenau-03} for a general discussion). E.g., in his treatment of
  the Hertz lattice, Nesterenko \cite{Nesterenko-83} used an expansion of
  $q_k$ and his compact solution thus slightly differs from the presented
  above. According to the discussion above, both solutions capture qualitative
  properties of the true lattice solution, but none of them can be expected to
  be exact in some asymptotic sense.  }

\subsection{Traveling waves in the lattice}
\label{sec:twla}

\begin{figure}
  \begin{center}
    \subfigure[]{\label{fig:twa_a}\includegraphics[draft=false,angle=270,width=0.45\textwidth]{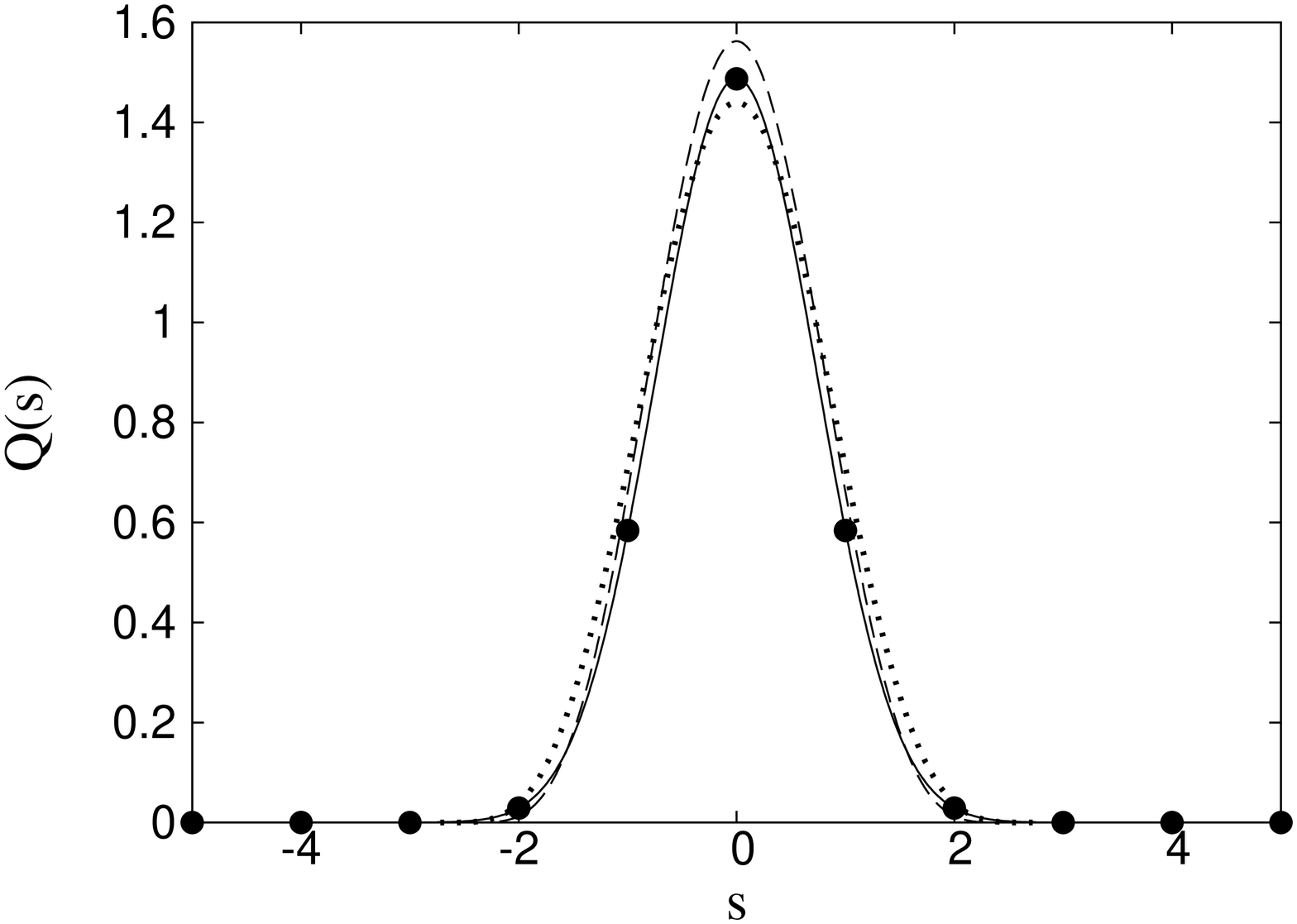}}
    \subfigure[]{\label{fig:twa_b}\includegraphics[draft=false,angle=270,width=0.45\textwidth]{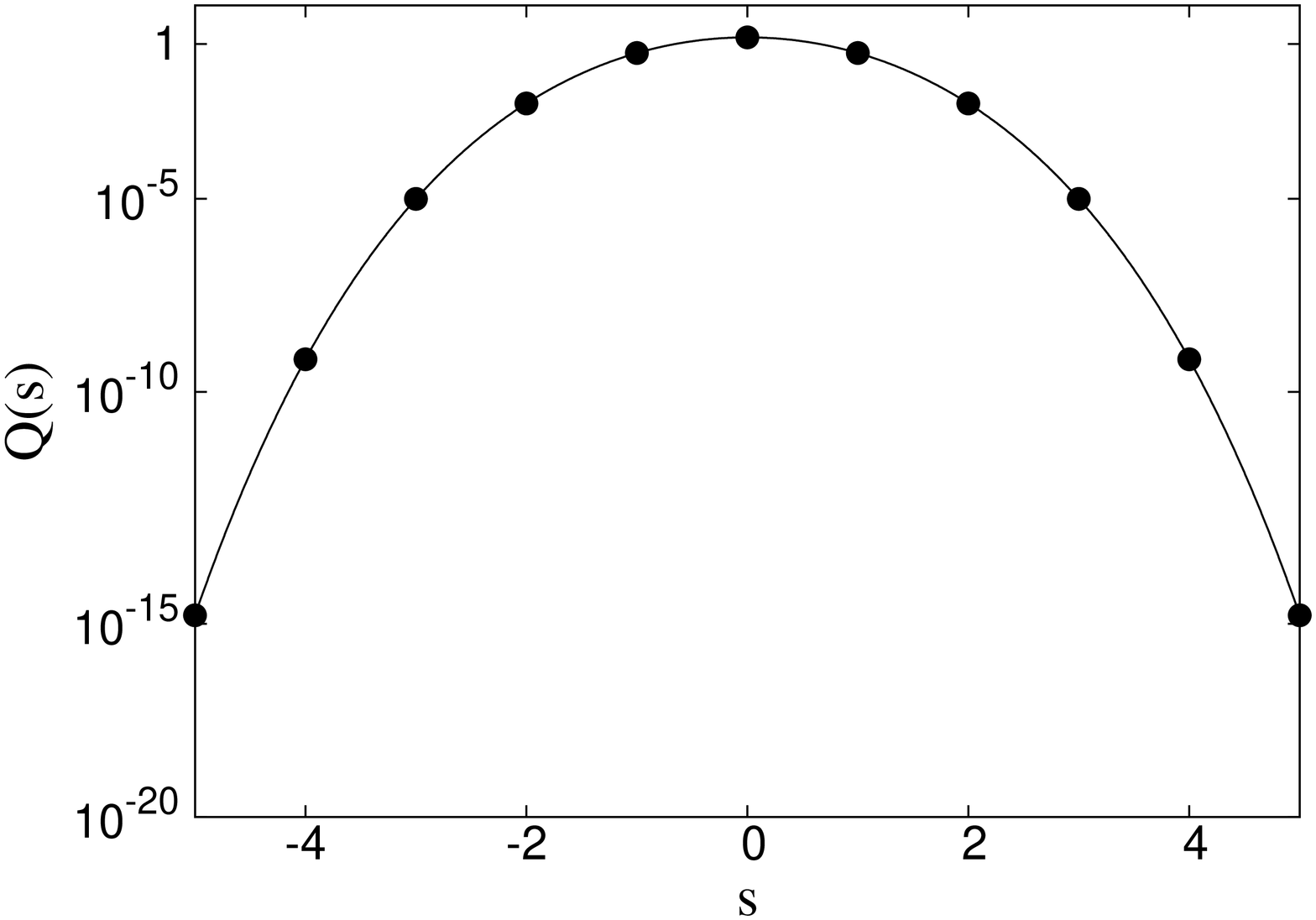}}
    \subfigure[]{\label{fig:twa_c}\includegraphics[draft=false,angle=270,width=0.45\textwidth]{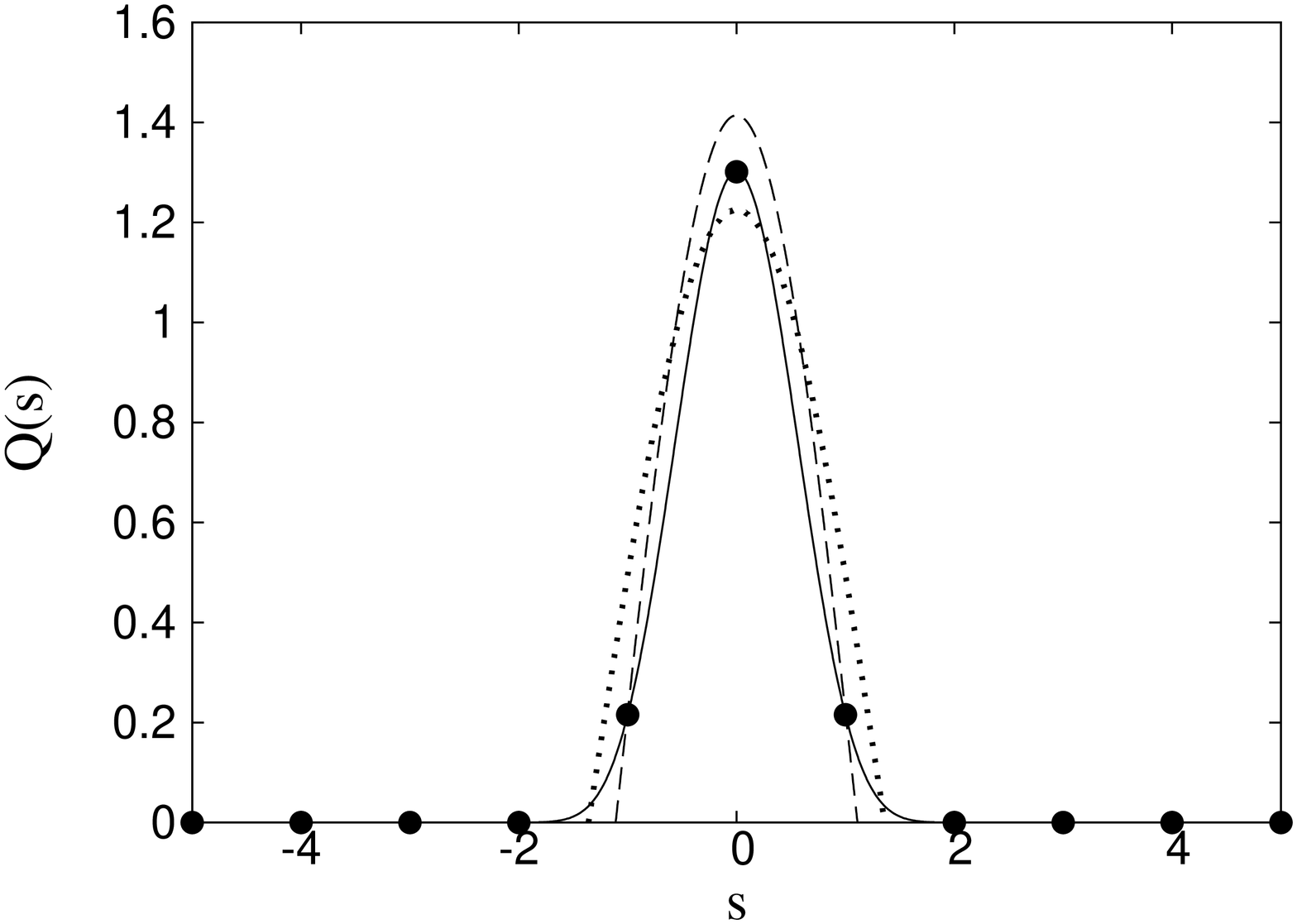}}
    \subfigure[]{\label{fig:twa_d}\includegraphics[draft=false,angle=270,width=0.45\textwidth]{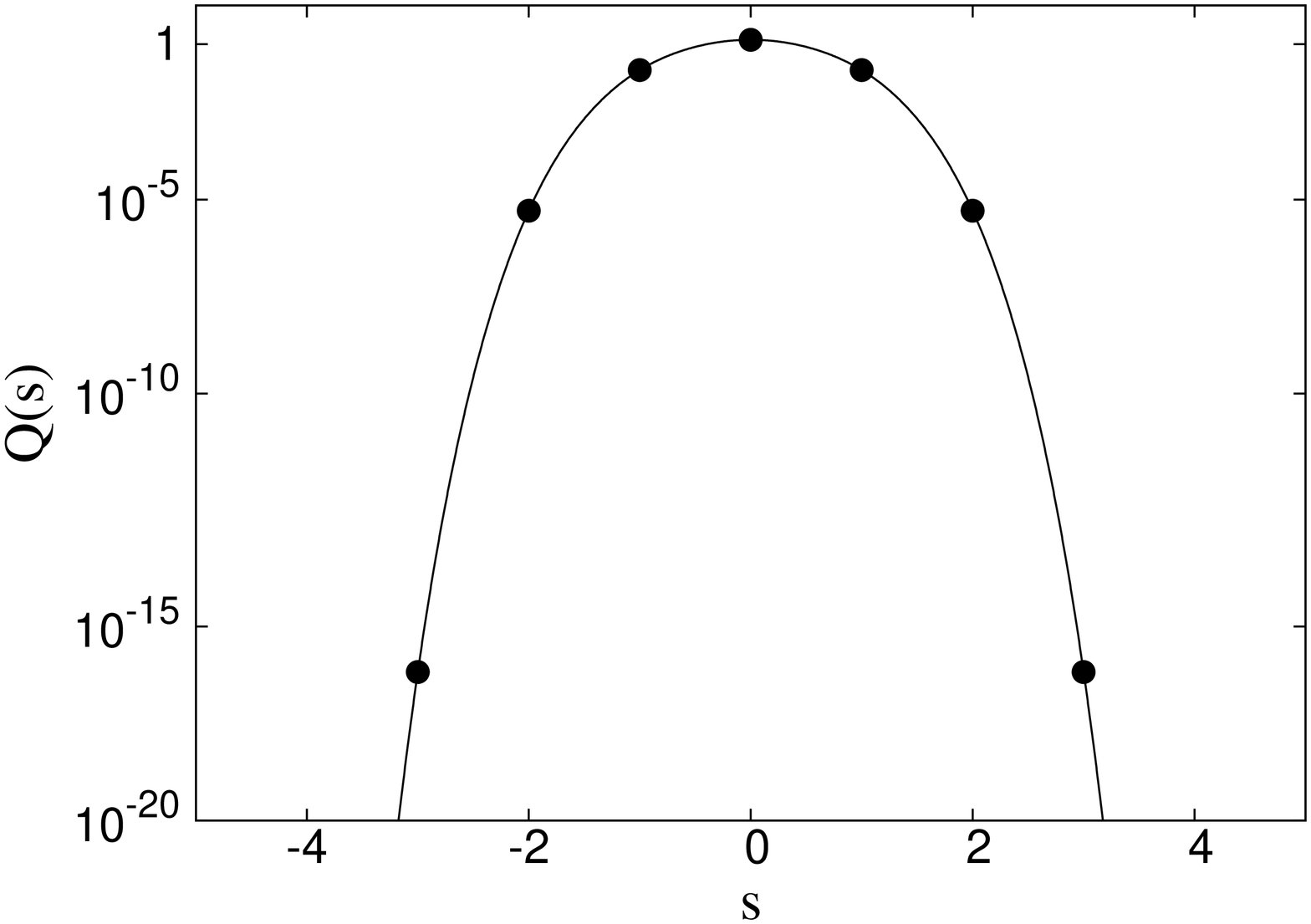}}
    \subfigure[]{\label{fig:twa_e}\includegraphics[draft=false,angle=270,width=0.45\textwidth]{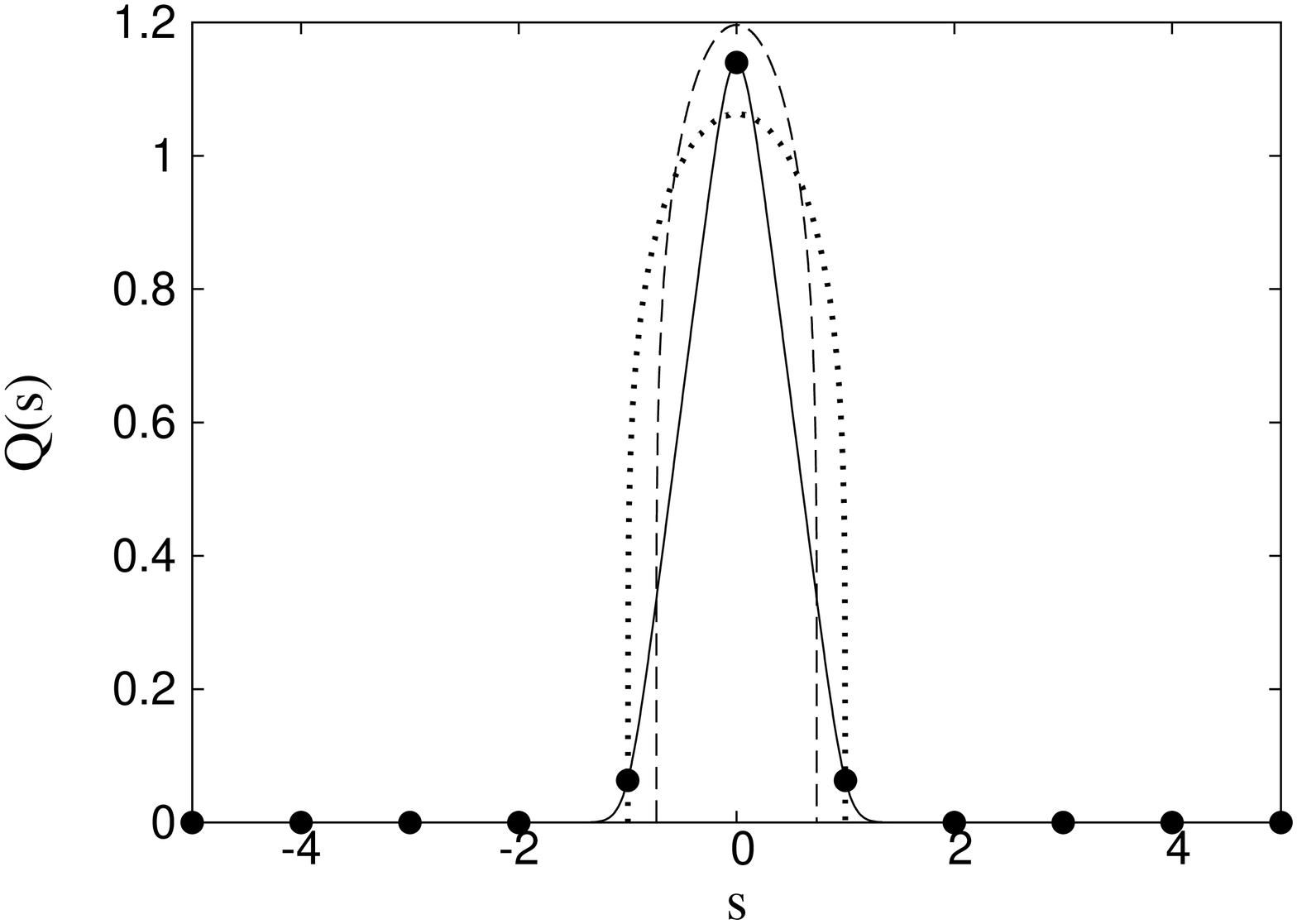}}
    \subfigure[]{\label{fig:twa_f}\includegraphics[draft=false,angle=270,width=0.45\textwidth]{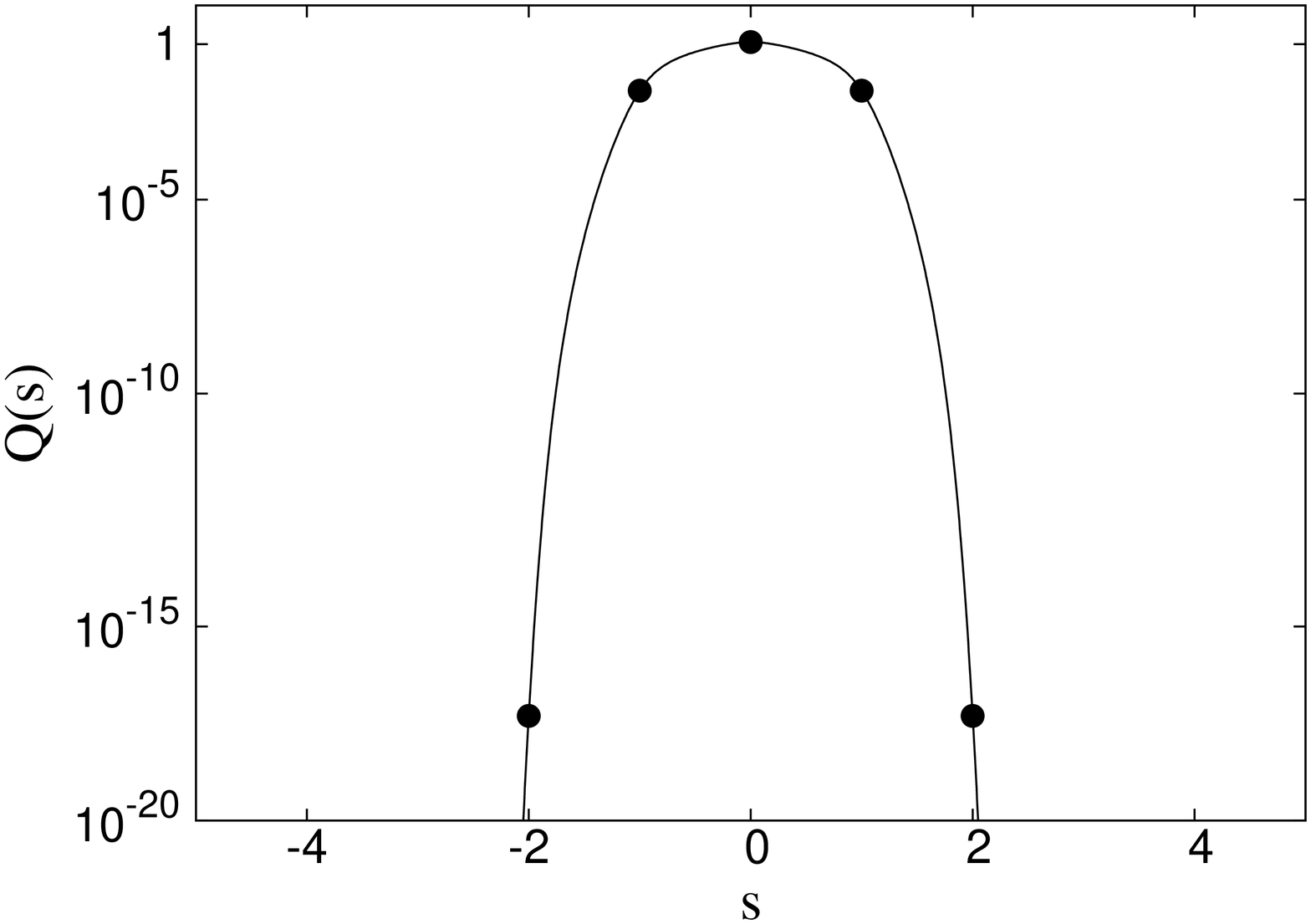}}
  \end{center}
  \caption{The traveling waves obtained from \eref{eq:integ_lattice_twa} for
    various powers $n$. Markers show the wave on the lattice, dotted lines
    show the corresponding solutions of the quasi-continuous approximation
    \eref{eq:qca} and dashed lines show the solutions of the QCA
    \eref{eq:qca_direct}. Left column: normal scale, right column: logarithmic
    scale. (a,b) $n=3/2$; (c,d) $n=3$, (e,f) $n=11$. Note that the width $w$
    of the compacton decreases as $n$ increases.}
  \label{fig:twa}
\end{figure}

In the lattice, the traveling wave ansatz reads $Q_k(t)=Q(k-\lambda
t)=Q(s)$. Inserting this ansatz into the lattice equations
\reff{eq:motion_diff} yields
\begin{equation}
\lambda^2 Q''(s) = Q^n(s-1) - 2 Q^n(s) + Q^n(s+1) \textrm{.}
\label{eq:lattice_twa}
\end{equation}
We employ now the scaling \reff{eq:resc} and set $\lambda=1$.  As demonstrated
in~\cite{Treschev-04}, this advanced-delay differential equation can be
equivalently written as an integral equation \be Q(s) = \int_{s-1}^{s+1}
(1-|s-\xi |) Q^n(\xi) \de \xi\;.  \ee{eq:integ_lattice} [One can easily check
  the equivalence by differentiating \reff{eq:integ_lattice} twice.]  We can
now, following the approach of
V. Petviashvili~\cite{Petviashvili-76,Petviashvili-81}, construct an iterative
numerical scheme to solve the integral equation
\eref{eq:integ_lattice}. Starting with some initial guess $Q_0$, one
constructs the next iteration via
\begin{subequations}
\label{eq:integ_lattice_twa}
\begin{eqnarray}
Q_{i+1} & = & \left(\frac{|| Q_i ||}{||Q_*||}\right)^\alpha Q_*
 \quad \textrm{and} \\
Q_{*} & = & \int_{s-1}^{s+1} (1-|s-\xi |) Q_i^n(\xi)\de\xi
\textrm{,}
\end{eqnarray}
\end{subequations}
(practically, we used the $L_1$-norm for $||\cdot||$). We have used
$\alpha=\frac{n}{n-1}$ that ensured convergence of the iterative scheme. The
integral in \eref{eq:integ_lattice_twa} was numerically approximated by virtue
of a $4$th-order Lagrangian integration scheme~\cite{Abramowitz-Stegun-64}. In
Fig.~\ref{fig:twa} the traveling waves for various powers $n$ are shown. Using
the logarithmic scale one clearly recognizes the compact nature of the waves.

\begin{figure}
  \begin{center}
    \includegraphics[draft=false,angle=270,width=0.8\textwidth]{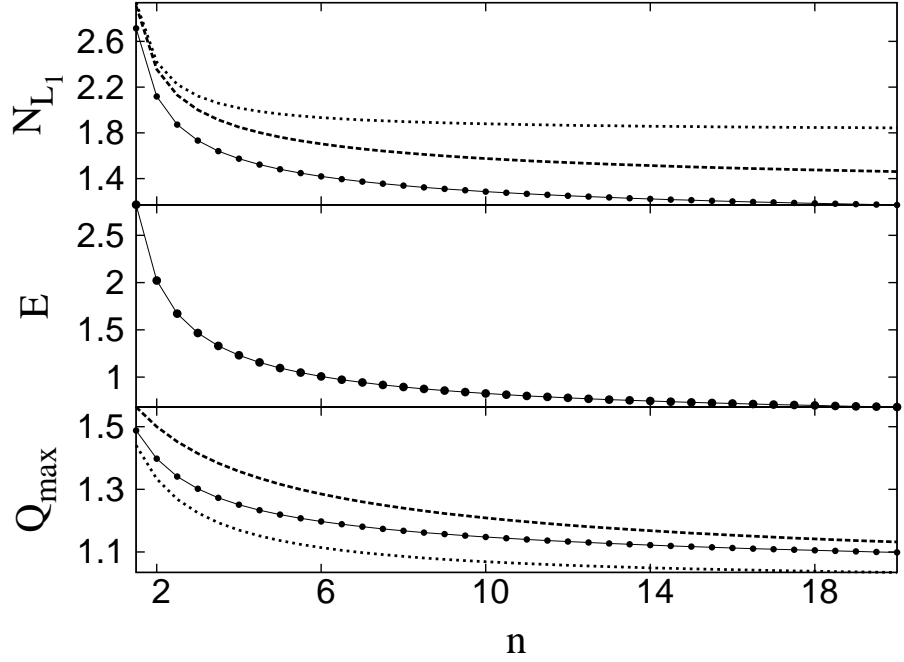}
  \end{center}
  \caption{ The dependency of the amplitude $Q_{max}$, the energy $E$ and the
    $L_1$-norm $N_{L_1}$ of a compacton on the nonlinearity index $n$. In this
    plots $\lambda=1$.  For comparison, the curves from the quasicontinuous
    approximation are shown with dotted lines for the Eq.~\eref{eq:qca} 
    and with dashed lines for Eq.~\eref{eq:qca_direct}.}
  \label{fig:twa_2_b}
\end{figure}

In Fig.~\ref{fig:twa_2_b} we show the dependencies of the total energy $E$,
the solution $L_1$-norm $N_{L_1}$, and the amplitude $Q_{max}$ of the found
waves on the nonlinearity index $n$, for a fixed wave velocity of
$\lambda=1$. Remarkably, the effective width $N_{L_1}/Q_{max}$ decreases with
increasing nonlinearity index and it seems that the profile of the compacton
converges to a triangular shape as $n\to\infty$.  A similar result has
  been obtained in \cite{Rosas-Lindenberg-04}, where the dependence of the
  pulse velocity on the nonlinearity index has been analyzed for large $n$ in the binary collision approximation.

\subsection{Estimation of the tails}

It is clear from the integral form \reff{eq:integ_lattice}, that the solution
cannot have a compact support.  In this section we estimate the decay of the
tails. We start with \eref{eq:integ_lattice} and substitute $Q(s)=e^{-f(s)}$:
\begin{equation}
Q(s) = \int\limits_{s-1}^{s+1} (1-|s-\xi |) e^{-n f(\xi)} \de \xi
\label{eq:est_tails_start}
\text{.} 
\end{equation}
We consider the tail for large $s>0$, if we assume a rapid decay of $Q(s)$,
then the integrand in \reff{eq:est_tails_start} has a sharp maximum at
$s-1$. Thus we can approximate the integral using the Laplace method.  At the
maximum we expand $f(\xi)$ into a Taylor series around $s-1$, keeping only the
leading first-order term:
\begin{equation}
Q(s) \approx \int\limits_{s-1}^{s+1} (1-|s-\xi |)
\exp \big\{ -n f(s-1) - n f'(s-1) (\xi-(s-1)) \big\} \de \xi
\end{equation}
We shift the integration range
\begin{equation}
Q(s)\approx e^{- n f(s-1)} \int\limits_0^2 \xi
e^{ - n f'(s-1) \xi} \de \xi
\end{equation}
where we also replace the decreasing part of the kernel with $\xi$. Since this
integrand decreases very fast, we can set the upper bound of the integration
to infinity, then by partial integration we obtain \be Q(s)=e^{-f(s)} \approx
\frac{ e^{-n f(s-1)}}{\big[ n f'(s-1) \big]^2} \text{.}\ee{eq:pint} Taking the
logarithm of this equation yields
\begin{equation}
- f(s) = -n f(s-1) - 2 \log[ n f'(s-1)]
\end{equation}
Since we expect that $f(s)$ is a rapidly growing function of $s$, we can
neglect the logarithmic term and obtain
\begin{equation}
f(s) = nf(s-1)
\end{equation}
This equation is solved by 
\begin{equation}
f(s) = C n^s = C e^{\log (n) s}
\end{equation}
where $C$ is an arbitrary constant. Finally we obtain
that the tail decays super-exponentially:
\begin{equation}
Q(s) = e^{-f(s)} \approx e^{-Cn^s}= \exp\Big[ -  C \exp \big( \log(n) s \big) \Big]
\label{eq:tails_estimate}
\end{equation}
This expression was first obtained by Chatterjee~\cite{Chatterjee-99} using
a direct expansion of the advanced-delayed equation~\reff{eq:lattice_twa}.

\begin{figure}
  \begin{center}
    \subfigure[]{
      \label{fig:tails_a}
     \includegraphics[draft=false,angle=270,width=0.45\textwidth]{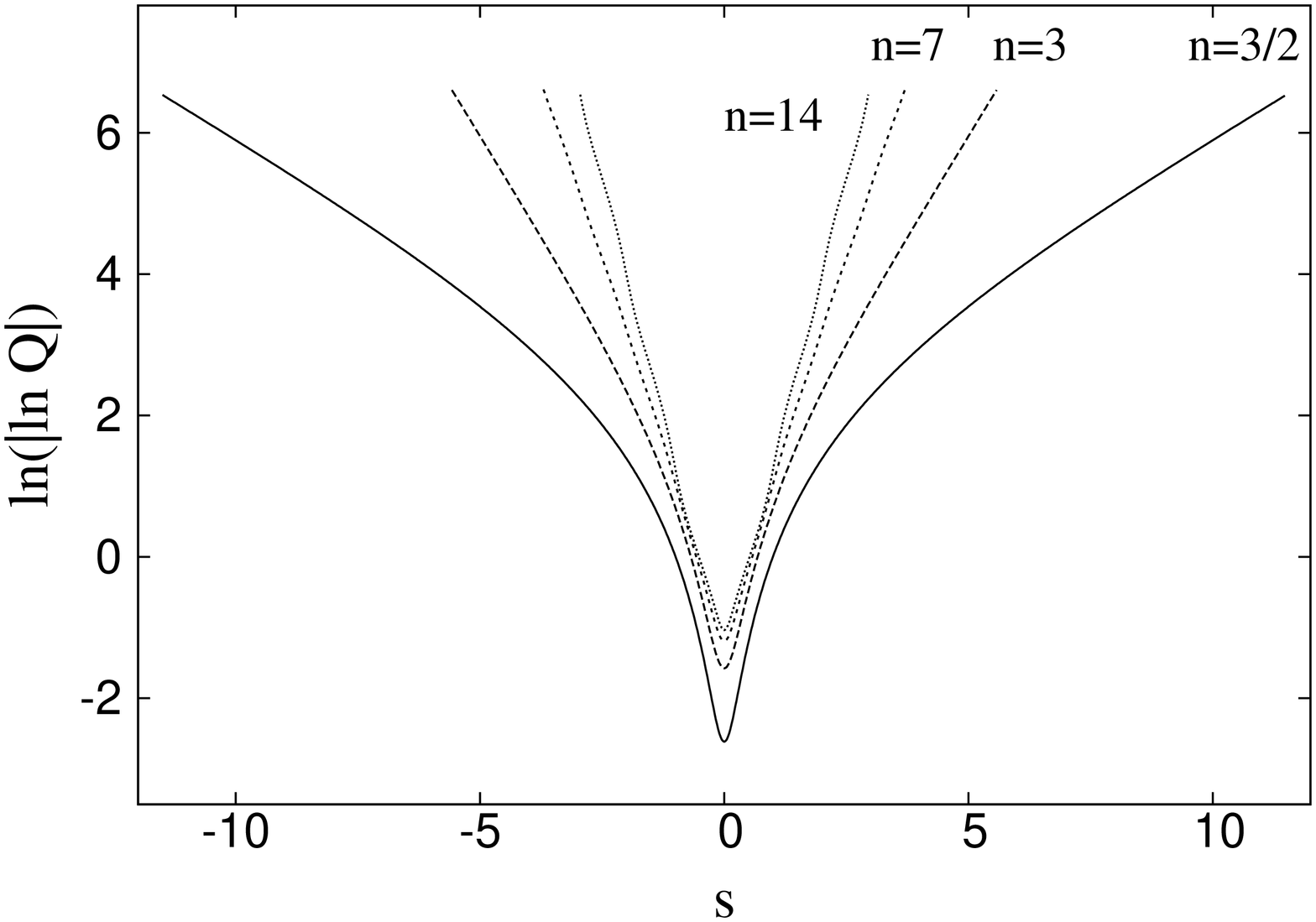}
    }
    \subfigure[]{
      \label{fig:tails_b}
      \includegraphics[draft=false,angle=270,width=0.45\textwidth]{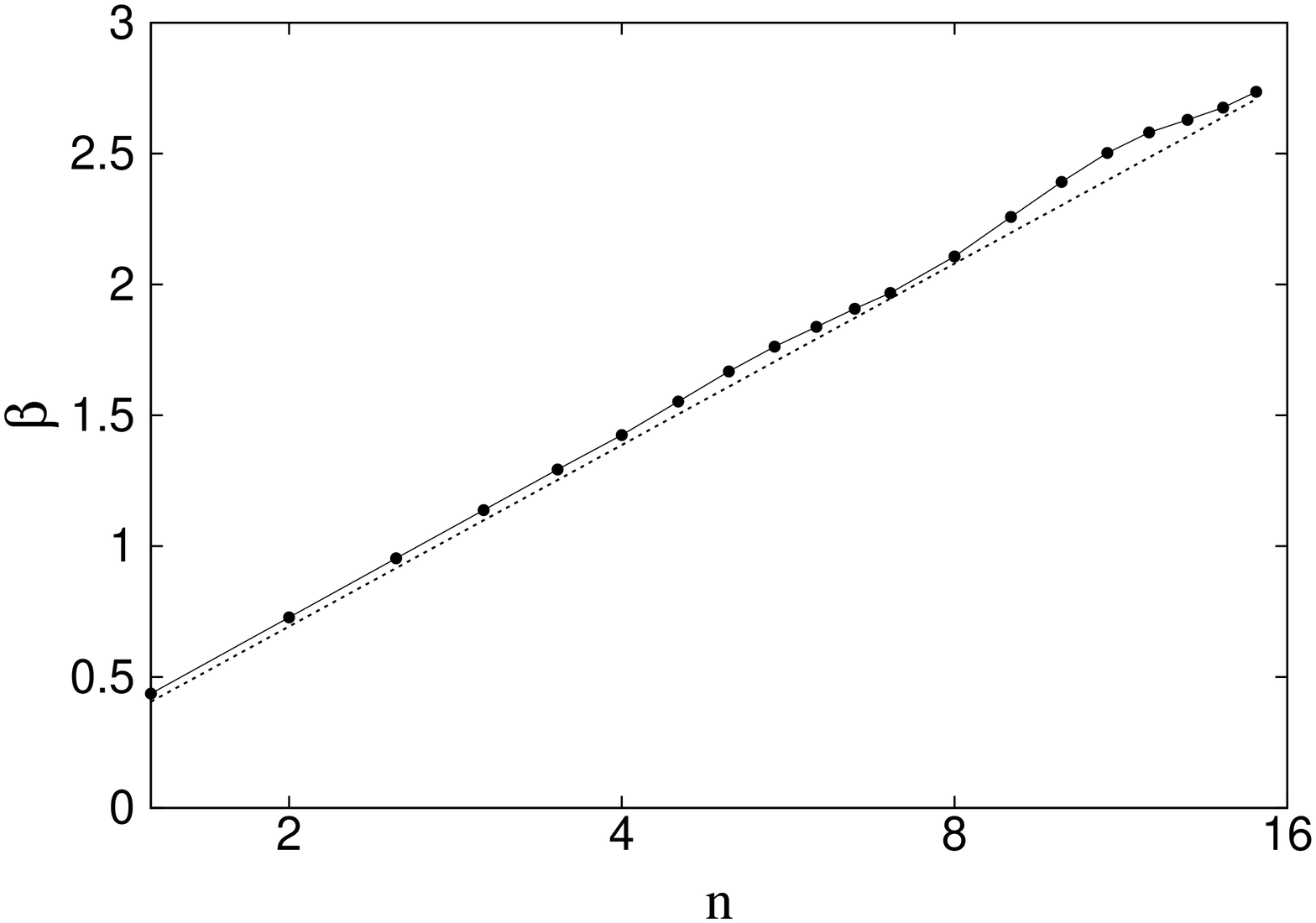}
    }
  \end{center}
  \caption{(a) The tails of the compactons in double logarithmic scale. (b)
    Comparison of the estimate \eref{eq:tails_estimate} with the compactons
    obtained from \eref{eq:integ_lattice_twa}.}
  \label{fig:tails}
\end{figure}

In Fig.~\ref{fig:tails_a} we show the tails of the compactons for various
values of $n$ and in Fig.~\ref{fig:tails_b} we compare the estimated decay
rate \eref{eq:tails_estimate} with compactons obtained numerically from the
traveling wave scheme \eref{eq:integ_lattice_twa}. To obtain the double
logarithmic decay rate $\beta=d\log(|\log(Q(s))|)/ds$, we first  compute
$\log(|\log(Q(s))|)$ and then the derivate is calculated using a spline smoothing
scheme \cite{Ahnert-Abel-07}. To suppress small oscillations of the tails we
average the numerical obtained derivative in the last $1/6$ of the compacton
domain. The numerical value of $\beta$ is shown in Fig.~\ref{fig:tails_b}.
Both coincide very well.

\section{Evolution and collisions of compactons}
\label{sec:evcol}

\subsection{Appearance of compactons from localized initial conditions}

\begin{figure}
  \begin{center}
    \subfigure[]{\label{fig:step_evolution_a}
      \includegraphics[draft=false,angle=270,width=0.45\textwidth]{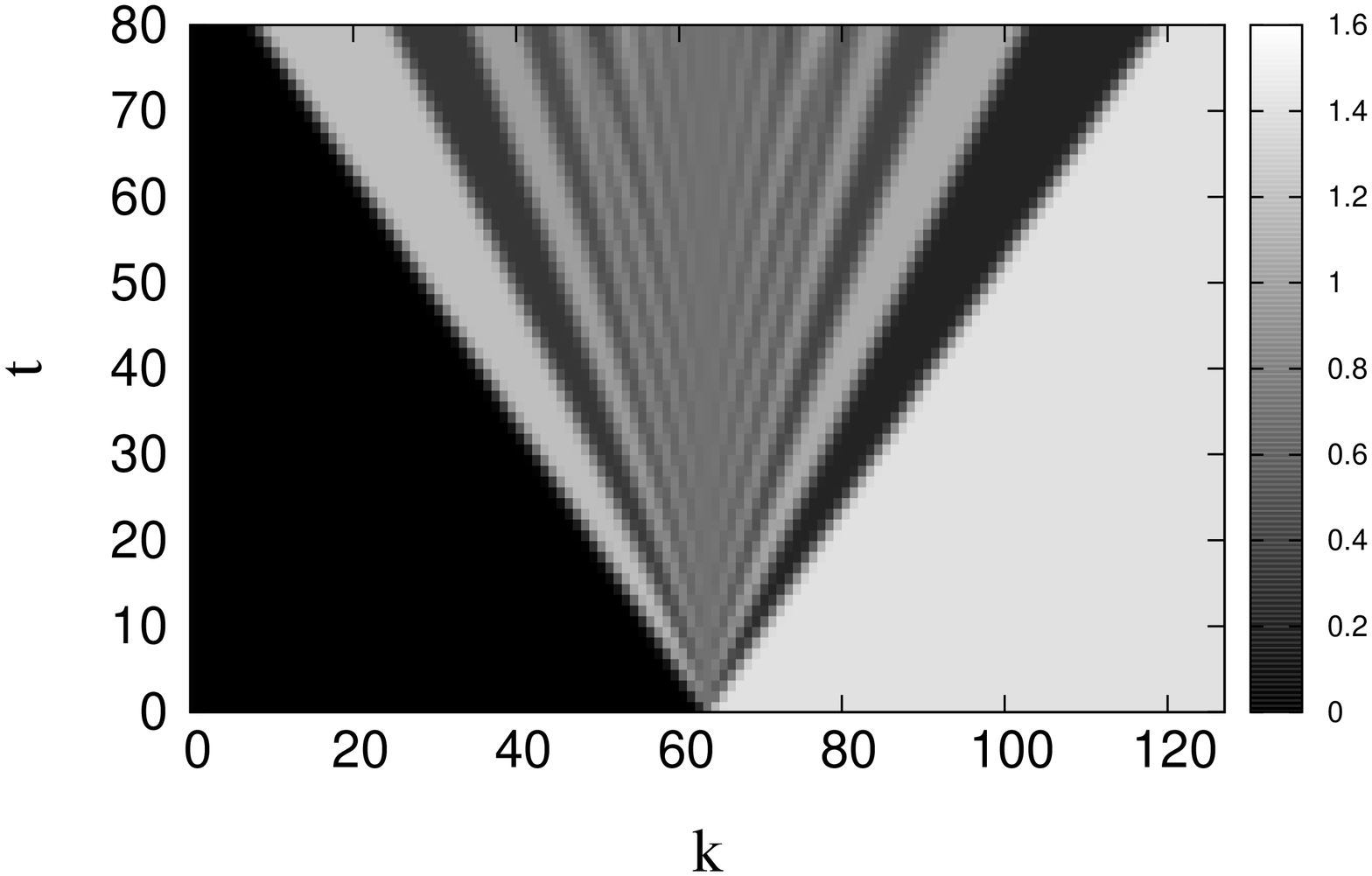}
    }
    \subfigure[]{\label{fig:step_evolution_b}
      \includegraphics[draft=false,angle=270,width=0.45\textwidth]{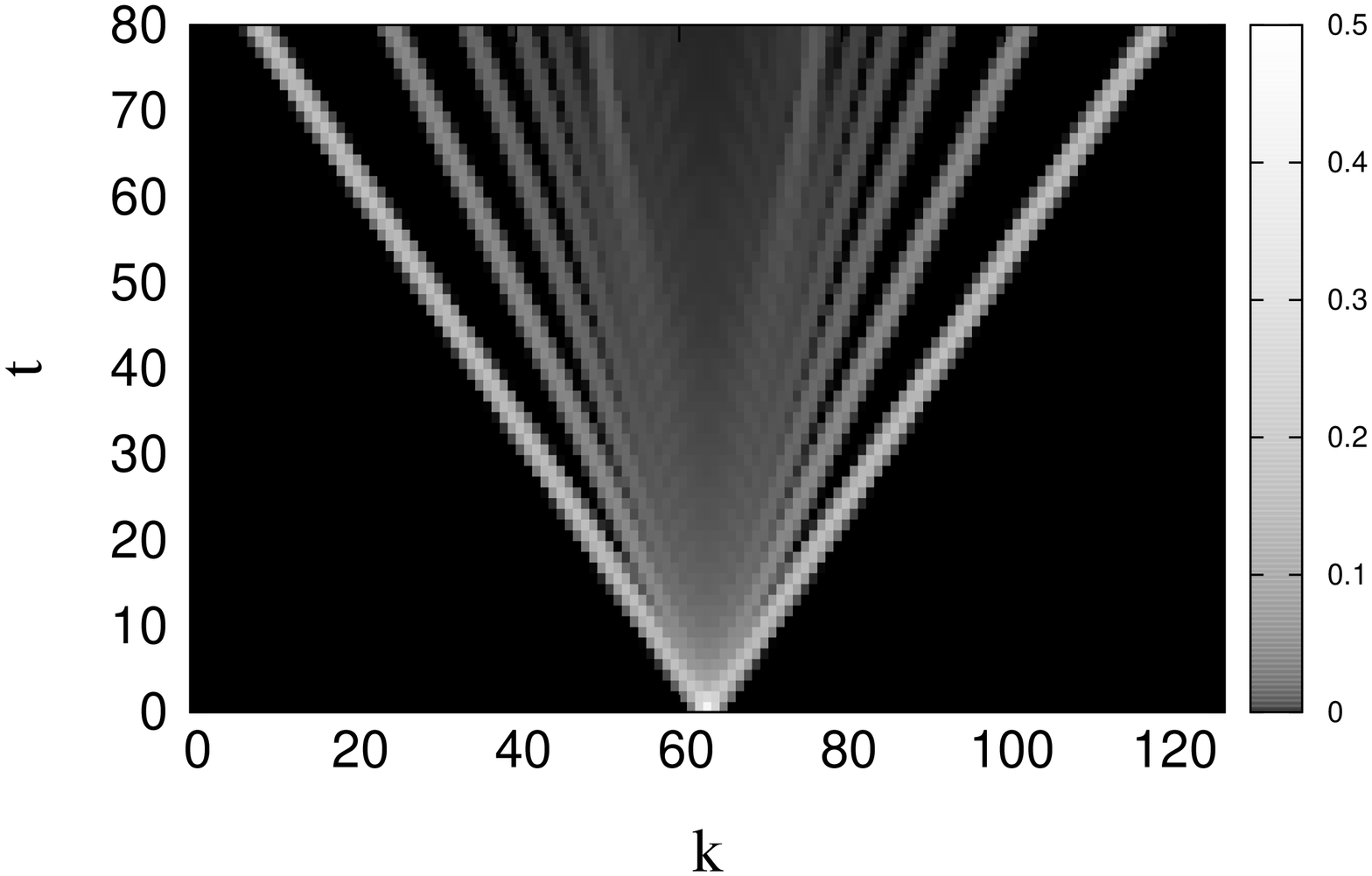}
    }
    \subfigure[]{\label{fig:step_evolution_c}
      \includegraphics[draft=false,angle=270,width=0.45\textwidth]{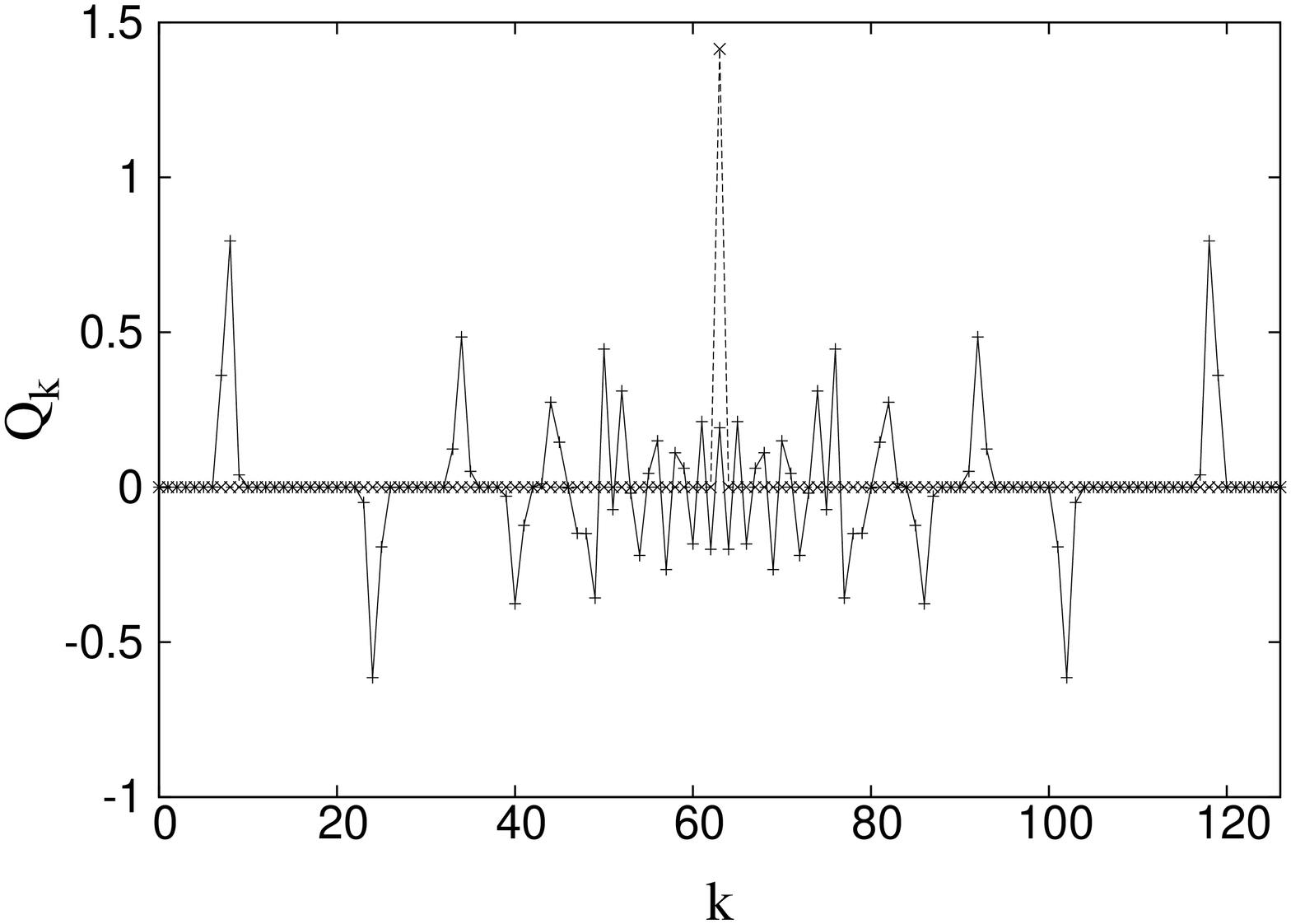}
    }
    \subfigure[]{\label{fig:step_evolution_d}
      \includegraphics[draft=false,angle=270,width=0.45\textwidth]{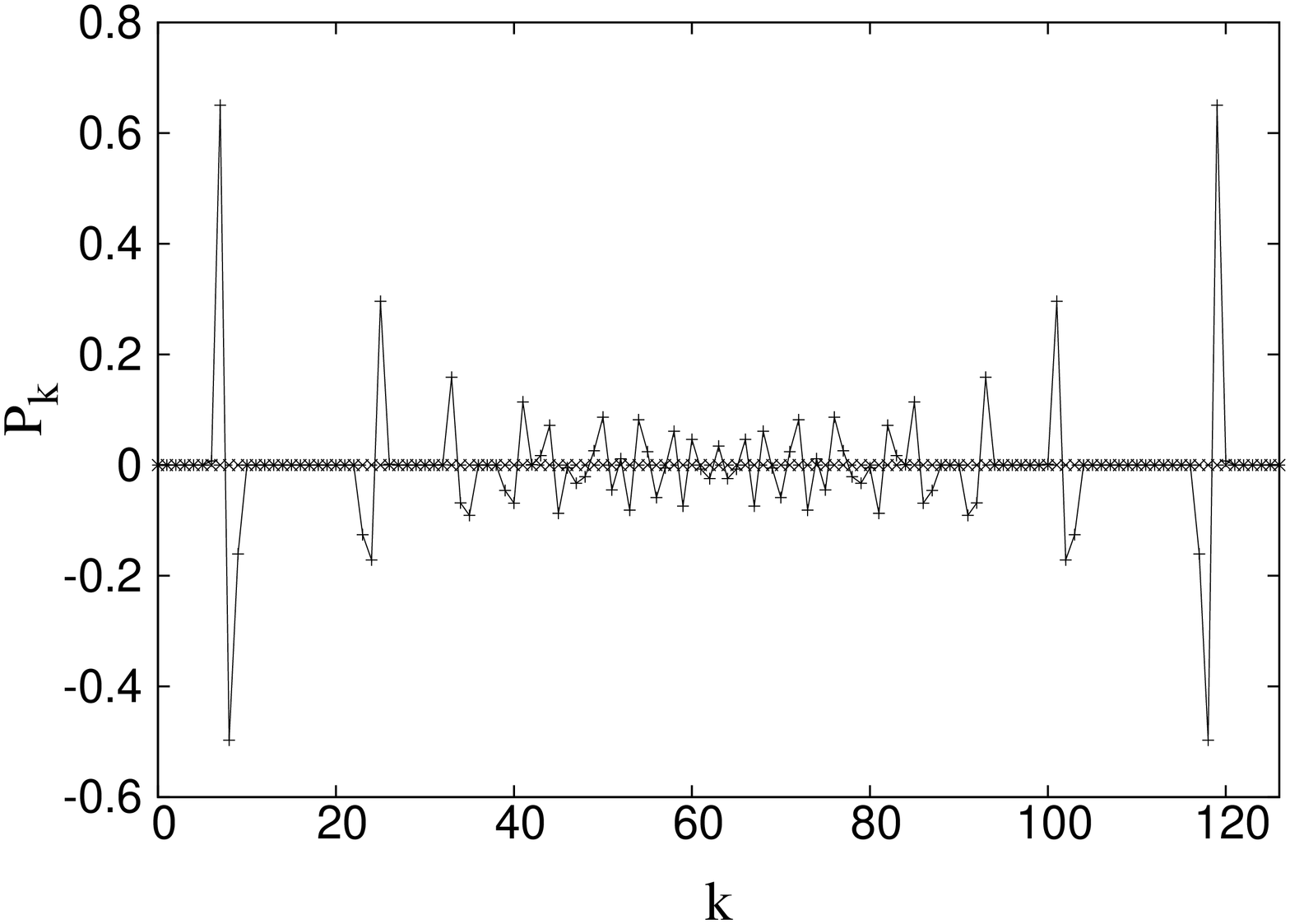}
    }
  \end{center}
  \caption{Evolution from an initial step for the nonlinearity index $n=3$,
    the lattice length is $N=128$ and open boundary conditions (hence
    $\ddot{q_1} = (q_2 - q_1)^n$ and $\ddot{q_N} = -(q_{N}-q_{N-1})^n$) are
    used. The initial conditions are $q_k(t=0) = (n+1)^{1/(n+1)}$ for $k>64$
    and $0$ else, initial momenta are zero.  Different plots show different
    quantities of the lattice: (a) the coordinates $q_k$; (b) the energy ${\cal E}_k$
    defined in \eref{eq:site_energy_definition}; (c) the difference
    coordinates $Q_k=q_{k+1}-q_k$ at time $t=80$, the initial state at $t=0$
    is shown here as the dashed line and (d) the difference momenta
    $P_k=p_{k+1}-p_k$ at $t=80$. The compactons originating from this initial
    state are clearly separated near the borders of the chain, those in the
    middle part are still overlap-ed.}
  \label{fig:step_evolution}
\end{figure}

\begin{figure}
  \begin{center}
    \subfigure[]{\label{fig:step_power_a}
      \includegraphics[draft=false,angle=270,width=0.31\textwidth]{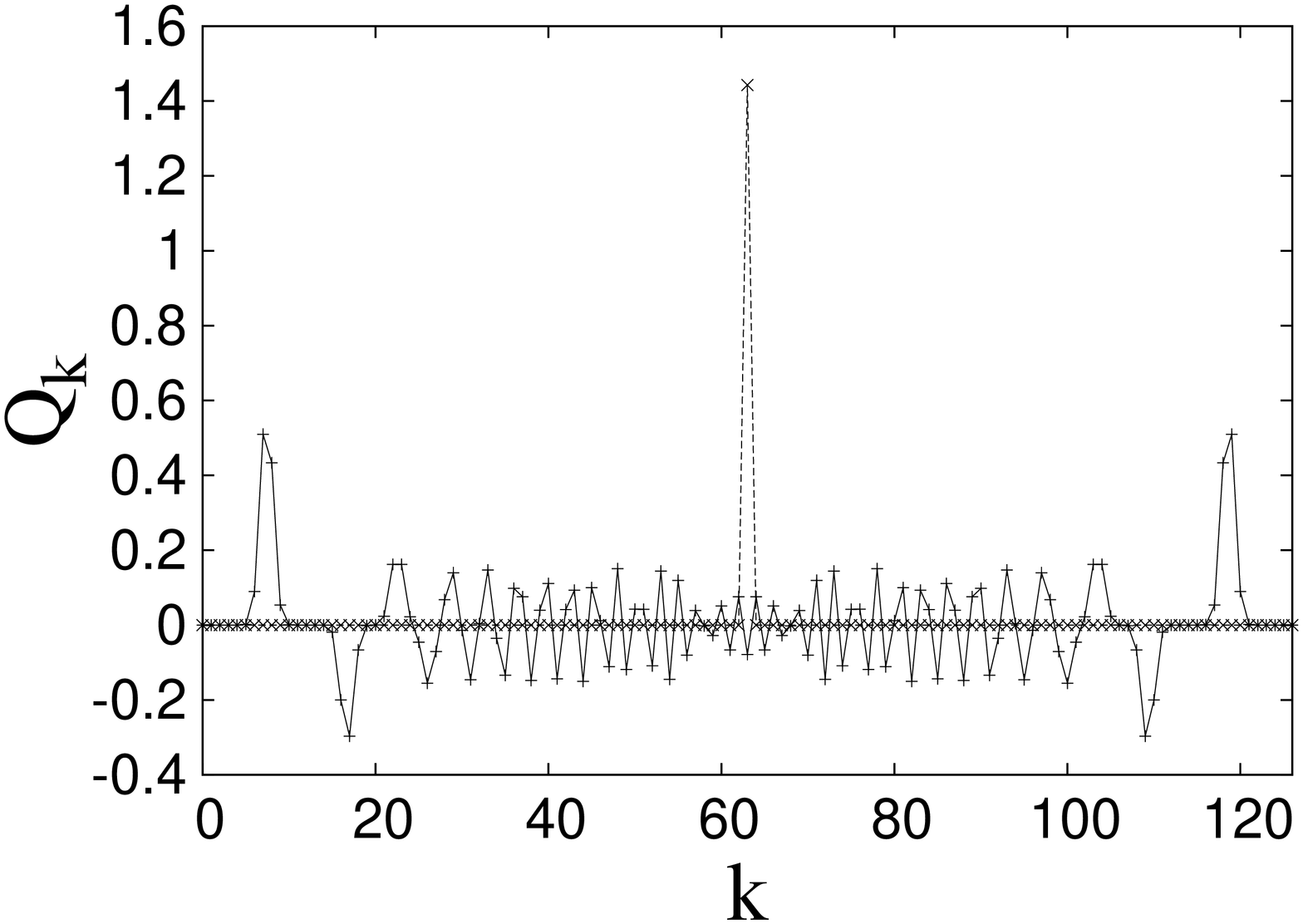}
    }
    \subfigure[]{\label{fig:step_power_b}
      \includegraphics[draft=false,angle=270,width=0.31\textwidth]{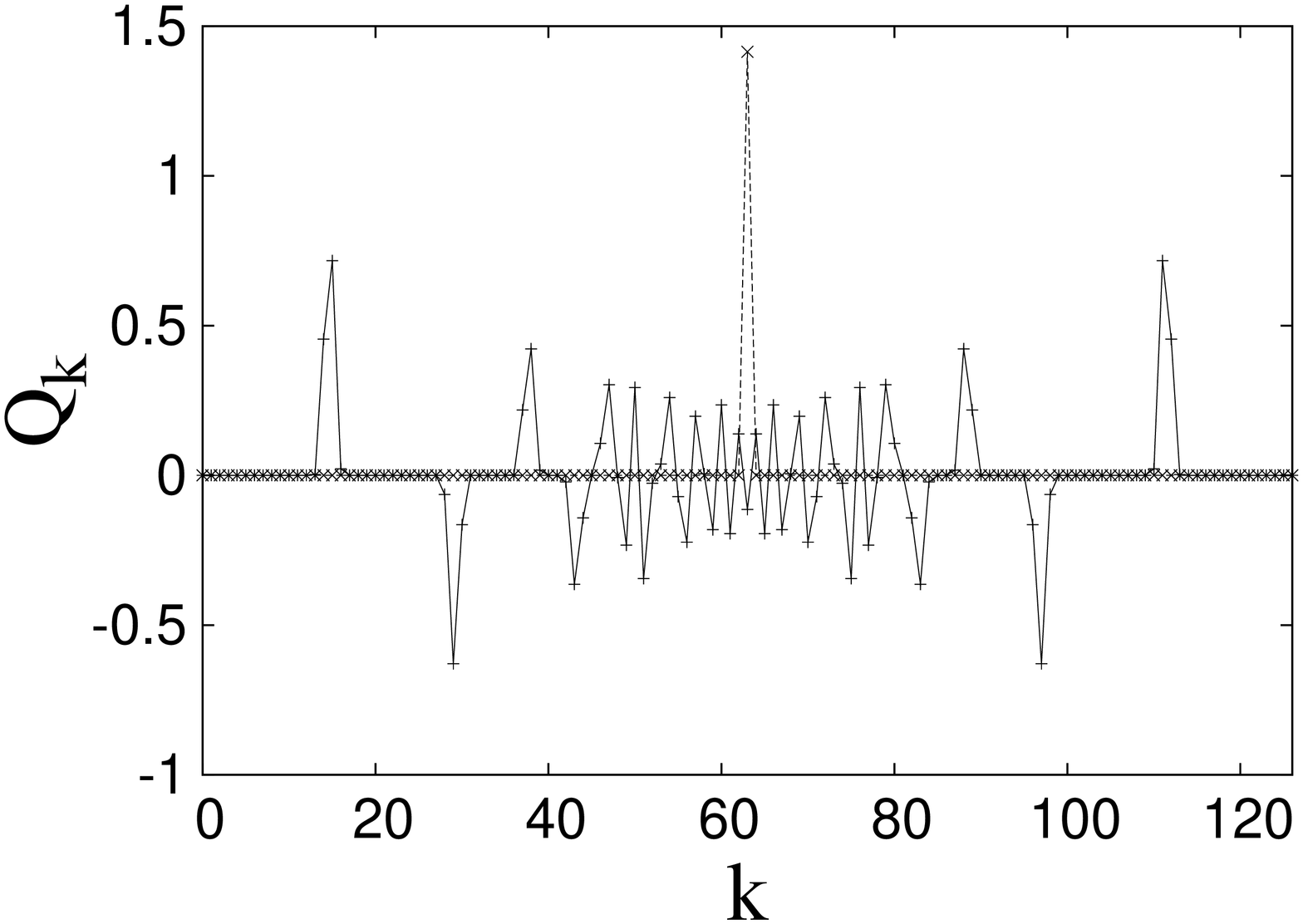}
    }
    \subfigure[]{\label{fig:step_power_c}
      \includegraphics[draft=false,angle=270,width=0.31\textwidth]{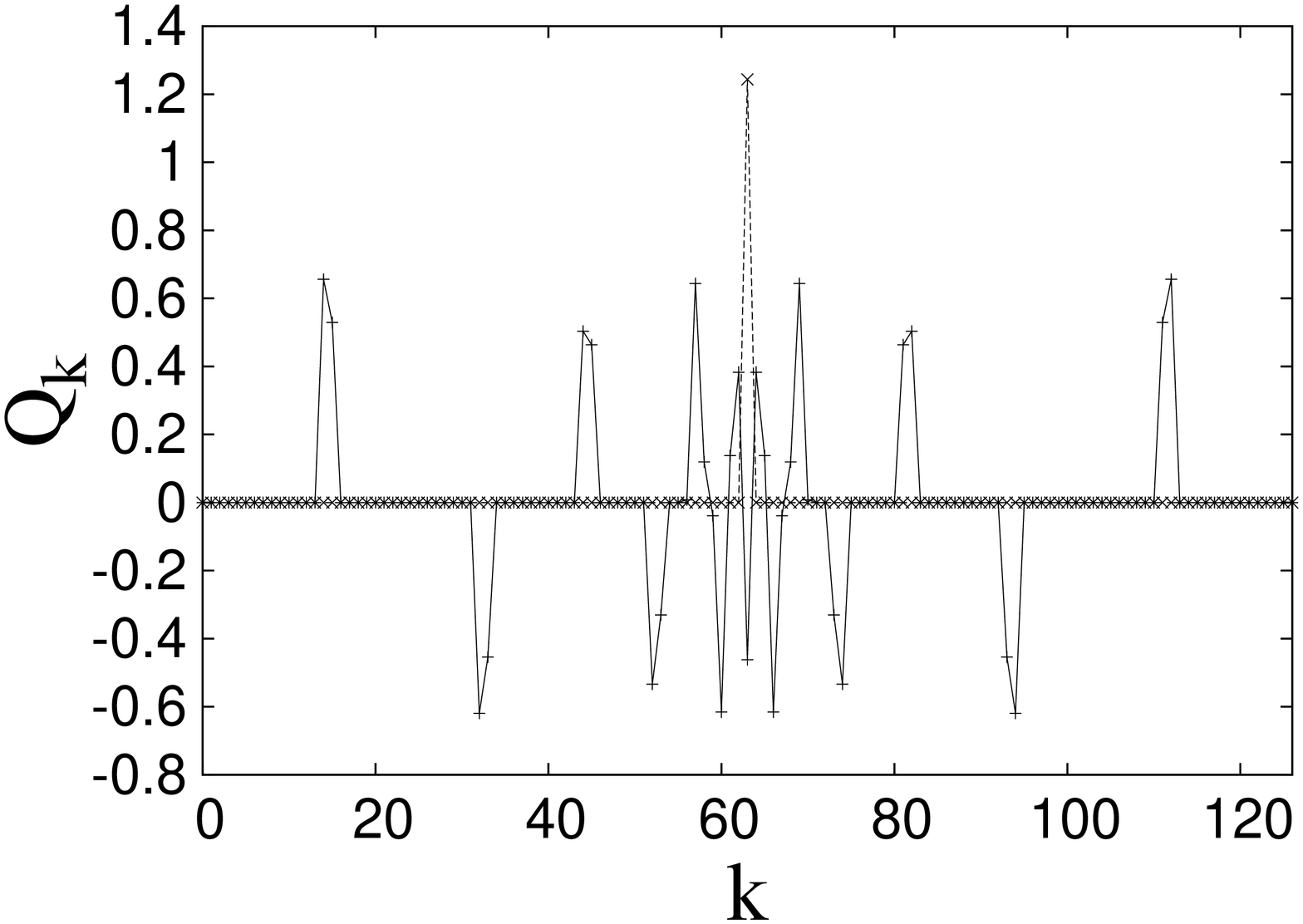}
    }
  \end{center}
  \caption{Evolution of an initial step (like in
    Fig.~\ref{fig:step_evolution}) for various nonlinearity indices (a)
    $n=1.5$, (b) $n=3$ and (c) $n=10$. The initial state is shown as dashed
    line, the solid line is the state at time $t=80$.}
  \label{fig:step_power}
\end{figure}

 The compact solitary waves constructed in the previous section are of
  relevance only if they evolve from rather general, physically realizable initial
  conditions. For an experimental significance (see~\cite{Lazaridi-Nesterenko-85,Coste-Falcon-Fauve-97} for experiments with Hertz beads) it is furthermore important, 
  that the emerging compact waves establish on
  relatively short distances, otherwise dissipation (which has not been
  considered here) will suppress their formation. We illustrate this in
Figs.~\ref{fig:step_evolution},\ref{fig:step_power}. There we report on a
numerical solution of the lattice equations \reff{eq:motion} on a finite
lattice of length $N=128$ (so that at the boundaries $\ddot{q_1} = |q_2 -
q_1|^n\sign(q_2 - q_1)$ and $\ddot{q_N} =
-|q_{N}-q_{N-1}|^n\sign(q_{N}-q_{N-1})$ holds). One of the quantities we
report is the local energy at site $k$ defined as
\begin{equation}
{\cal E}_k = \frac{p_k^2}{2} + \frac{1}{2(n+1)} \left( |q_{k+1}-q_k|^{n+1} + |q_k -
q_{k-1}|^{n+1} \right)
\text{.} \label{eq:site_energy_definition}
\end{equation}
As an initial condition we have chosen a kink in the variables $q_k$:
$q_k=(n+1)^{1/(n+1)}$ for $k>64$ and $q_k=0$ elsewhere. This profile has unit
energy, it corresponds to the localized initial condition in the variable $Q$:
$Q_{k}=\delta_{k,64}\cdot (n+1)^{1/(n+1)}$.  The evolution of different
variables is shown in Fig.~\ref{fig:step_evolution}.  From the initial pulse
of $Q$, a series of compactons with alternating signs is emitted in both
directions. The amplitude of the perturbation near the initially seeded site
decreases and correspondingly increases a characteristic time of the
evolution. We expect that at large times, compactons with small amplitudes
will continue to detach. In Fig.~\ref{fig:step_power} we show the evolution
from the initial step for different nonlinearities $n=1.5,\, 3,\,10$. The
plots look very similar and compactons are emitted in every case. The number
of the emitted compactons and their amplitudes depend on the nonlinearity
index.

\begin{figure}
  \begin{center}
    \subfigure[]{\label{fig:random_ic_a}
      \includegraphics[draft=false,angle=270,width=0.45\textwidth]{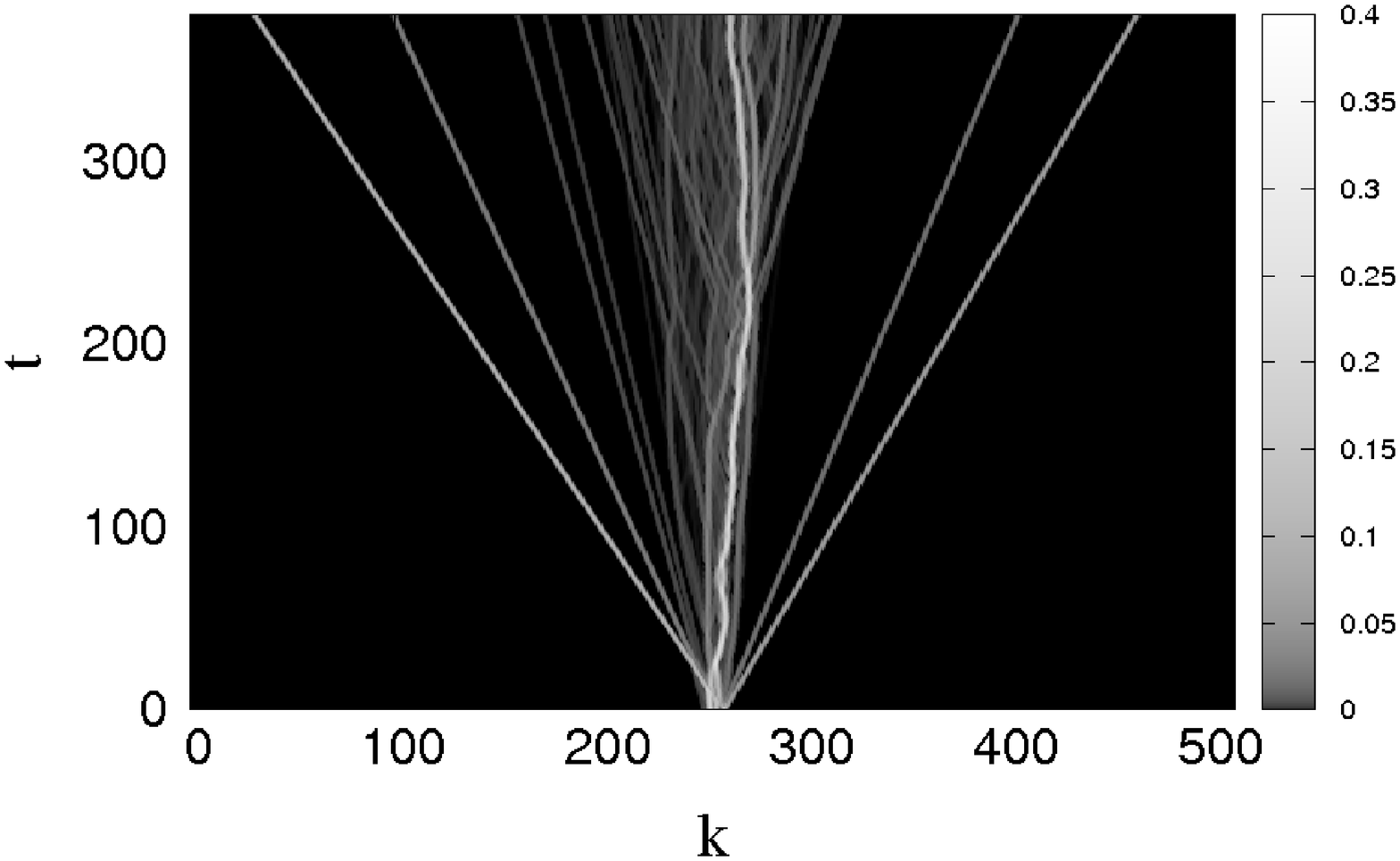}
    }
    \subfigure[]{\label{fig:random_ic_b}
      \includegraphics[draft=false,angle=270,width=0.45\textwidth]{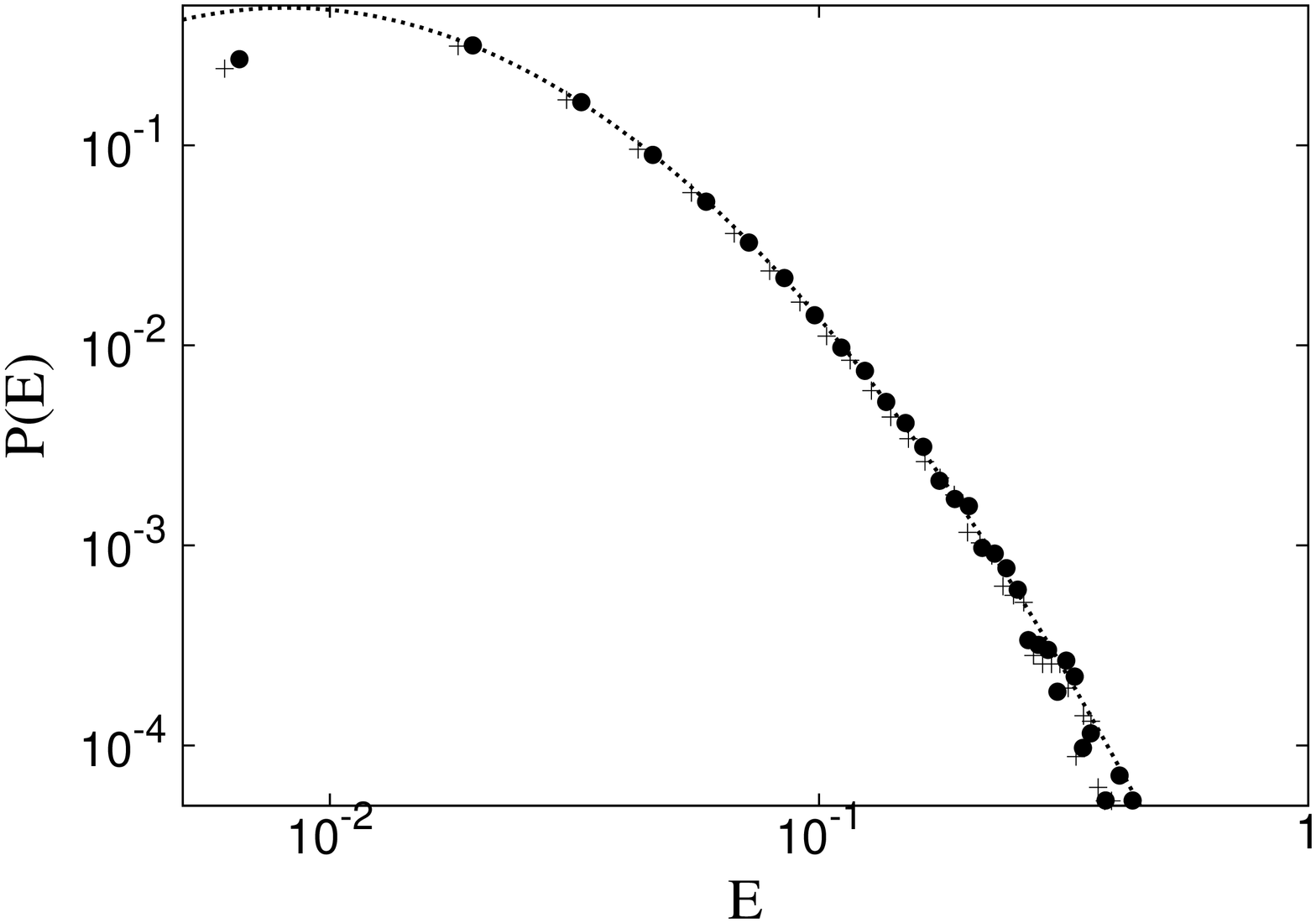}
    }
  \end{center}
  \caption{(a) Compactons emerging from localized random initial
    conditions. The nonlinearity index is $n=3$. The gray scale corresponds to
    the energy \eref{eq:site_energy_definition} of the lattice site. (b)
    Energy distribution of the compactons emitted from localized random
    initial conditions. The statistics was obtained from $60000$ simulations;
    in each simulation the lattice was integrated to the time $T=1000$ and the
    energy distribution of the compactons emerged to the right (black circles)
    and the left (crosses) have been determined. The distributions obeys in
    very good approximation $P(E) \sim E^{-a \log(E) - b}$, with $a=0.57$ and
    $b=5.47$.}
  \label{fig:random_ic}
\end{figure}

In our next numerical experiment we studied the emergence of compactons not
from a sharp step in the coordinates $q_k$, but from localized random initial
conditions. In Fig.~\ref{fig:random_ic_a} we show a typical evolution in a
lattice of length $N=512$ (with nonlinearity index $n=3$) resulting from
random initial conditions $q_k$ in the small region $N/2-5\leq k<N/2+5$ around
the center of the lattice. In this region the coordinates $q_k$ have been
chosen as independent random numbers, identically and symmetrically uniformly
distributed around zero, while $p_k(0)=0$. Furthermore, the energy of the
lattice was set to $E=1$ by rescaling. In a particular realization of
Fig.~\ref{fig:random_ic_a}, at the initial state two compactons emerge to
the right and four compactons to the left. In the center of the lattice a
chaotic region establishes and slowly spreads over the lattice, possibly
emitting more compactons on a longer time scale. In Fig.~\ref{fig:random_ic_b}
we perform a statistical analysis of this setup by showing the energy
distribution of compactons emitted from localized random initial conditions as
described above. This distribution was obtained from $60000$ simulations, in
each simulation the energy of the emitted compactons have been determined and
counted. The functional form of the distribution obeys in very good
approximation $P(E) \sim E^{-a \log(E) - b}$, with $a=0.57$ and $b=5.47$.

\subsection{Collisions of compactons}

\begin{figure}
  \begin{center}
    \subfigure[]{\label{fig:col_3.0_1}\includegraphics[draft=false,angle=270,width=0.45\textwidth]{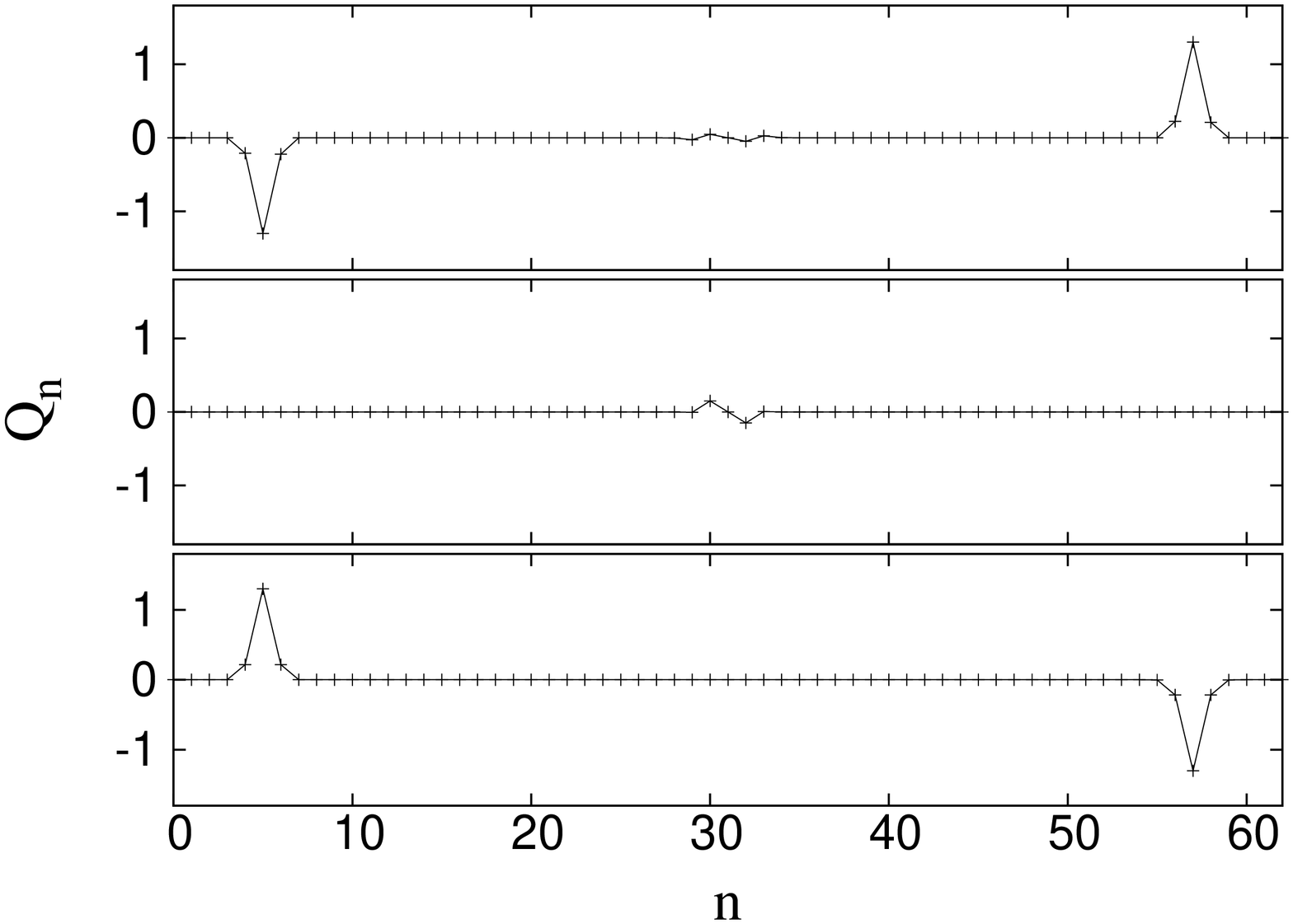}
    }
    \subfigure[]{\label{fig:col_3.0_2}\includegraphics[draft=false,angle=270,width=0.45\textwidth]{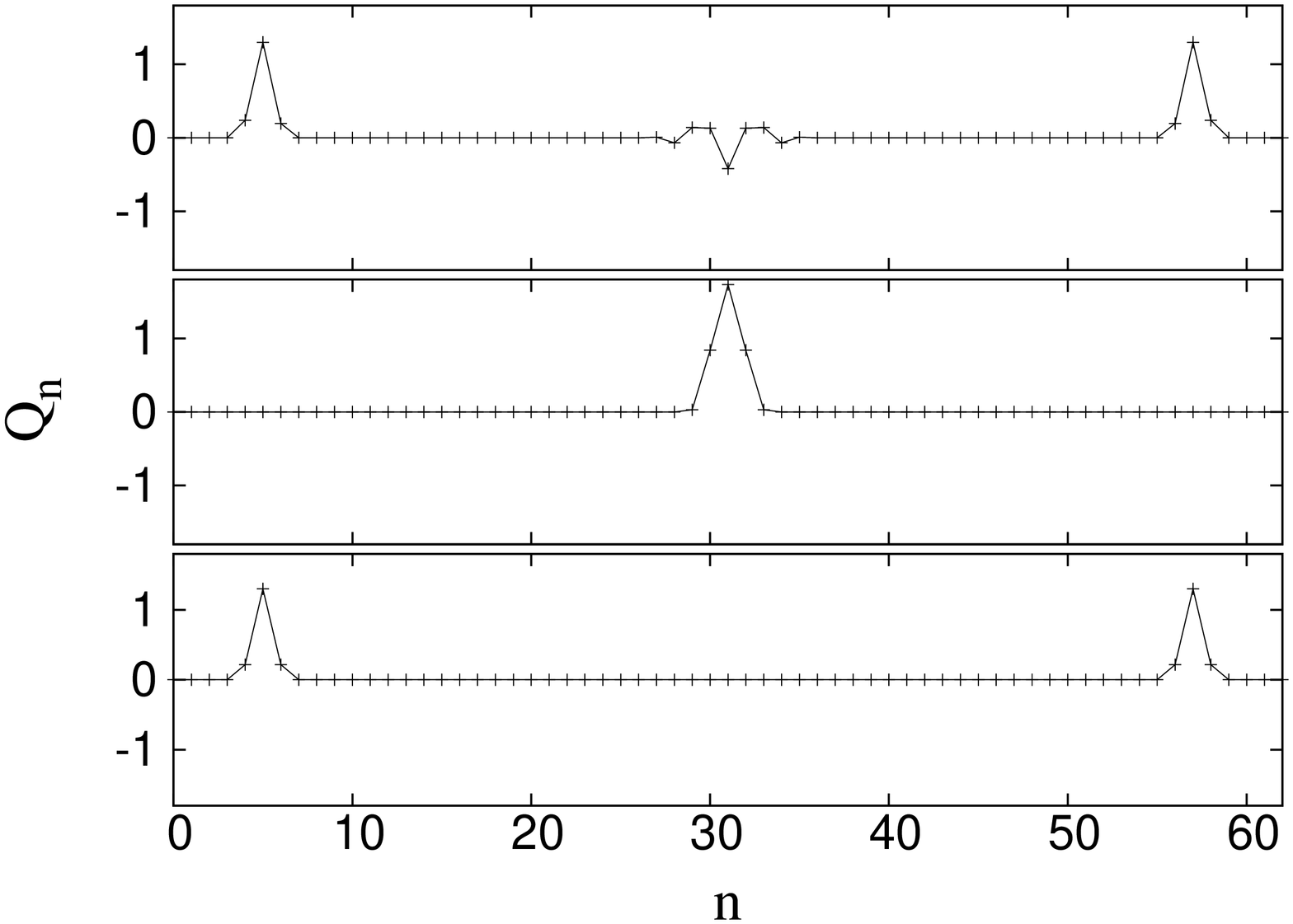}
    }
    \subfigure[]{\label{fig:col_3.0_3}\includegraphics[draft=false,angle=270,width=0.45\textwidth]{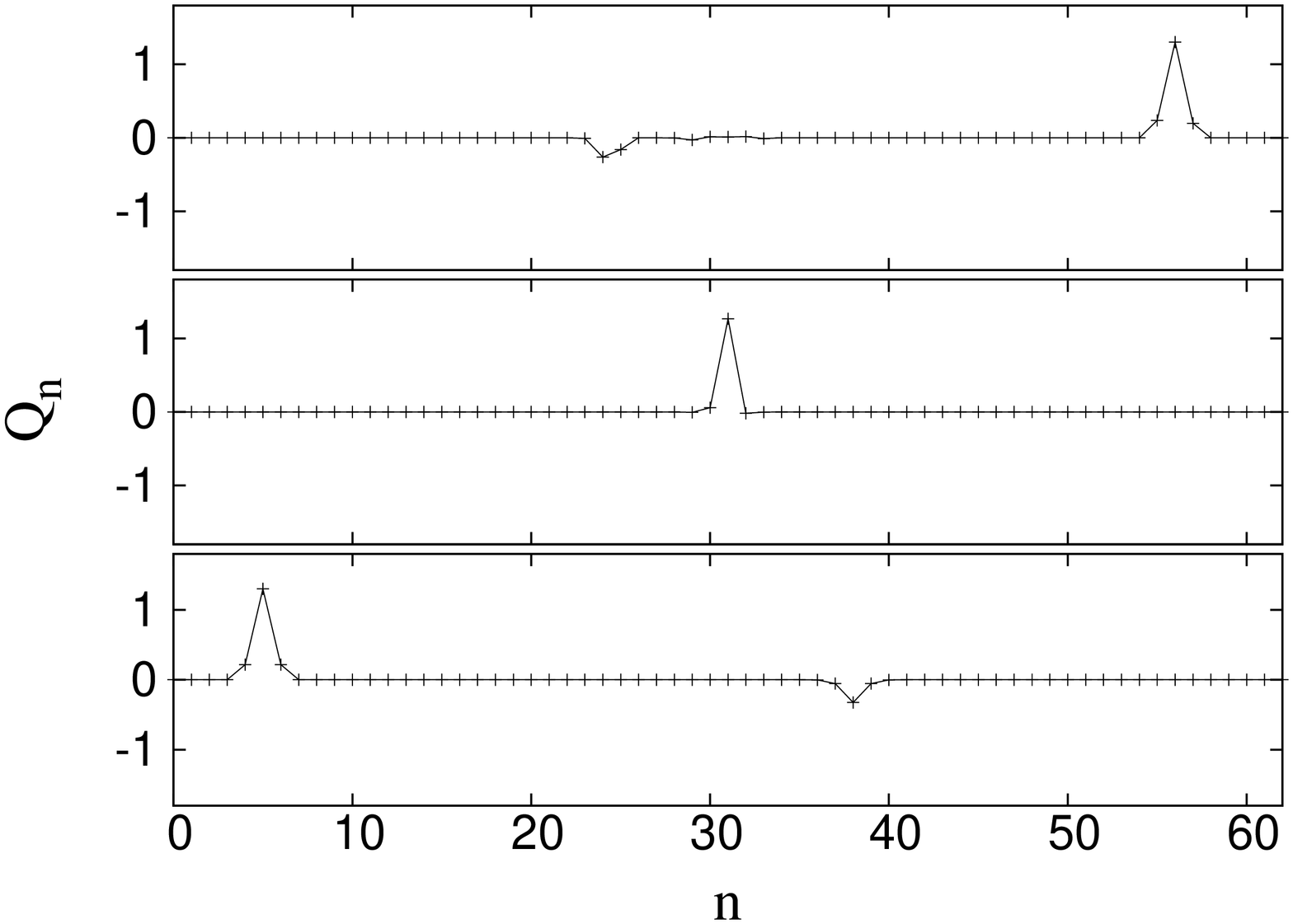}
    }
    \subfigure[]{\label{fig:col_3.0_4}\includegraphics[draft=false,angle=270,width=0.45\textwidth]{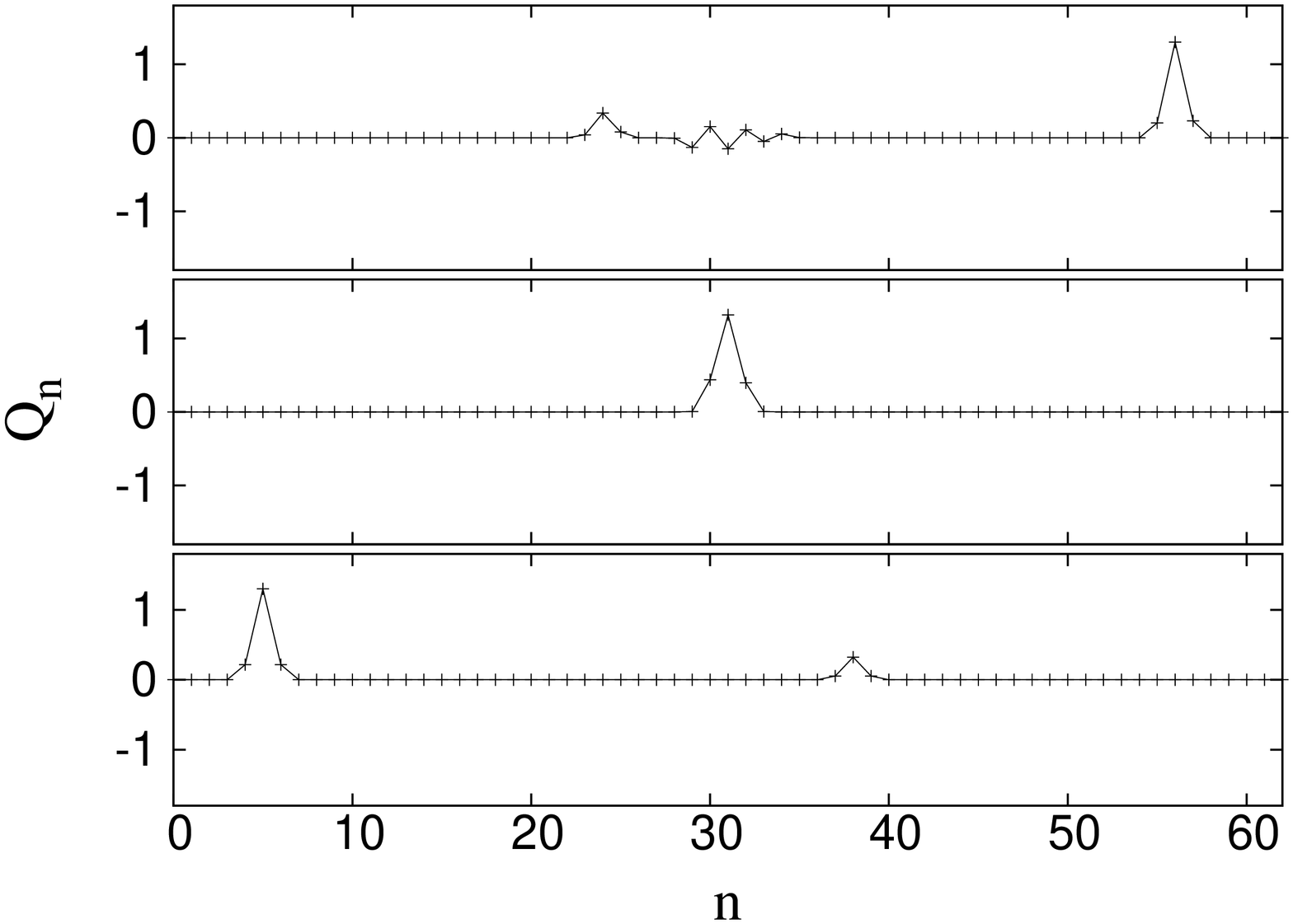}
    }
    \subfigure[]{\label{fig:col_3.0_5}\includegraphics[draft=false,angle=270,width=0.45\textwidth]{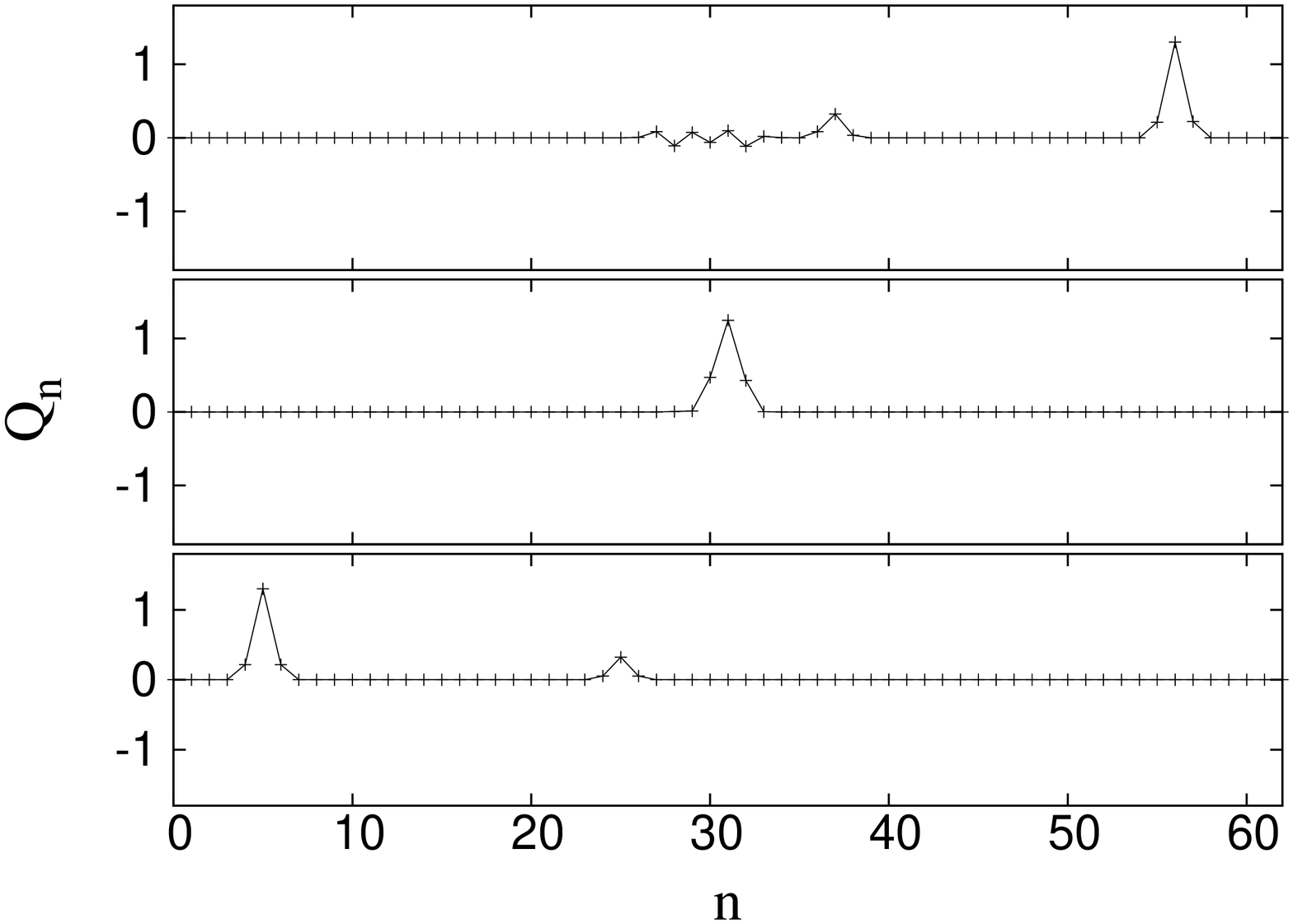}
    }
    \subfigure[]{\label{fig:col_3.0_6}\includegraphics[draft=false,angle=270,width=0.45\textwidth]{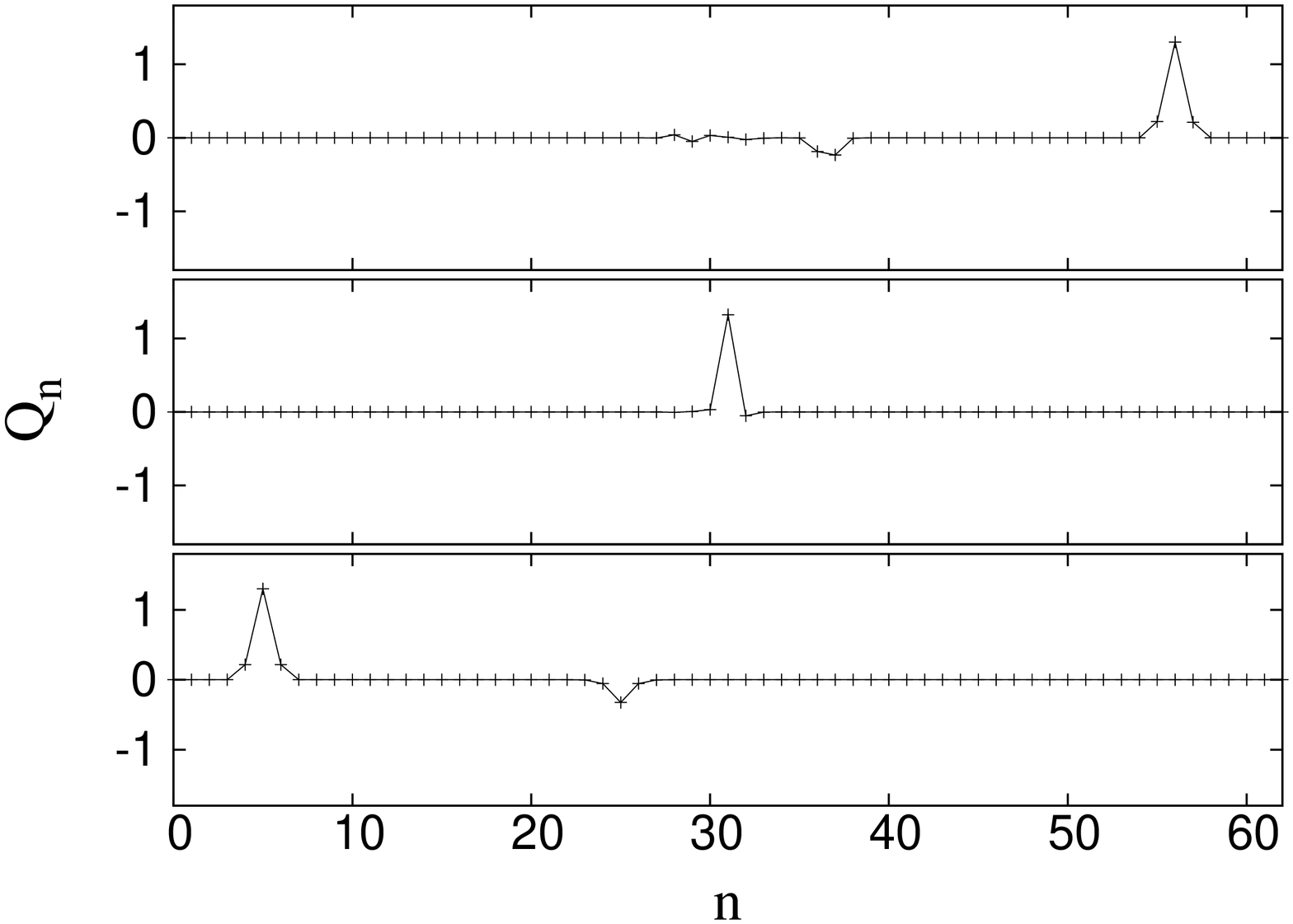}
    }
  \end{center}
  \caption{Collisions of compactons in the Hamiltonian lattice with $n=3$,
    shown are difference coordinates $Q_k$. We have considered 6 different
    collision scenarios, in each plot the lower panel is the initial
    configuration of the lattice, the middle panel is the state of the lattice
    at some time during the maximal overlap, and the upper panel shows the
    lattice past the collision. (a) Compactons of equal energy having opposite
    amplitudes and velocities; (b) compactons of equal energies and amplitudes
    but opposite velocities; (c) compactons of different energies having
    amplitudes and velocities of opposite signs; (d) compactons of different
    energies having amplitudes of the same and velocities of opposite signs;
    (e) compactons of different energies having velocities and amplitudes of
    the same sign; (f) compactons of different energies having velocities of
    the same sign and amplitudes of opposite signs. }
  \label{fig:col_3.0}
\end{figure}

As we have demonstrated above, compactons naturally appear from rather general
initial conditions. To characterize their stability during the evolution, we
study their stability to the collisions. This study is not complete but only
illustrative, as in Fig.~\ref{fig:col_3.0} we exemplify different cases of
collision of two compactons in a lattice with $n=3$. These six setups present
all possible scenarios of two compactons: (i) two colliding compactons with
the same amplitudes, (ii) two compactons with different amplitudes moving
toward each other, and (iii) two compactons with different amplitudes moving in
the same direction and passing each other. Each of these three cases has two
sub-cases, because the amplitudes can have the same or different sign. It
should be mentioned, that these six collisions do not represent the complete
picture of all collision, moreover we have not varied parameters such as the
distance between two colliding compactons or their amplitudes.

In all the cases presented, the initial compactons survive the collision: they
are not destroyed although they do not come out of the collision unchanged. In
all cases the collision is non-elastic, some small perturbations (that
presumably on a very long time scale may evolve into small-amplitude
compactons) appear.

\begin{figure}
  \begin{center}
    \subfigure[]{
      \label{fig:col_chaos_a}
      \includegraphics[draft=false,width=\textwidth]{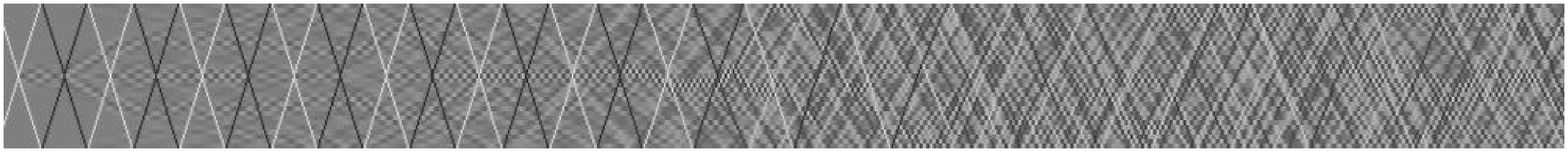}
    }
    \subfigure[]{
      \label{fig:col_chaos_b}
      \includegraphics[draft=false,width=\textwidth]{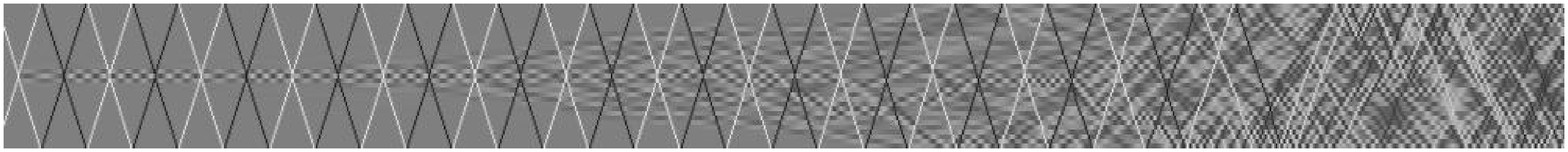}
    }
    \subfigure[]{
      \label{fig:col_chaos_d}
      \includegraphics[draft=false,width=\textwidth]{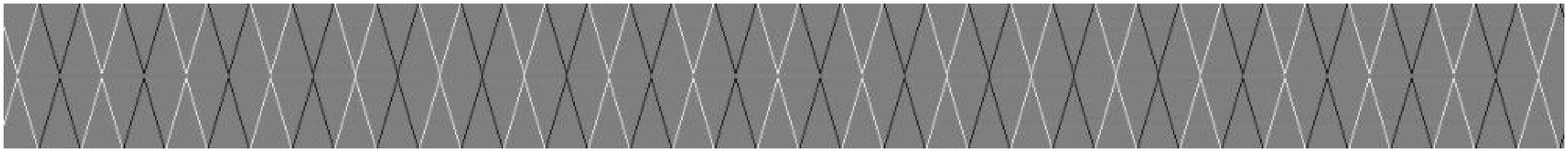}
    }
  \end{center}
  \caption{Collisions of compactons and emergence of chaos after multiple
    collisions. Different plots show different nonlinearity indices (a) $n=3$,
    (b) $n=9/2$ and (c) $n=11$. Time increases from left to right and the
    difference coordinates $Q_k$ are shown in gray scale. Remarkably the
    elasticity of the collision increases with increasing nonlinearity index
    $n$, so that practically no irregularity appears at $n>10$.}
  \label{fig:col_chaos}
\end{figure}

Because of this non-elasticity, on a finite lattice after multiple collisions
initial compactons get destroyed and a chaotic state appears in the lattice,
as illustrated in Fig.~\ref{fig:col_chaos}. There we show the evolution of the
two compactons with the same amplitude and sign of the amplitude for three
different nonlinearities: $n=3$, $n=9/2$ and $n=11$. In the first two cases
the chaotic state establishes relatively fast. In the third simulation with
$n=11$ the situation is different.  Here the chaotic state does not appear
even on a very long time scale. We run the simulation for very long times up
to $T=2\cdot 10^5$, but could not observe the development of a chaotic
state. We have checked this phenomenon also for higher values of $n$ with the
same result. Presumably, these initial conditions lie on a stable
quasiperiodic orbit or are extremely close to a such one.

\section{Chaos in a finite lattice}
\label{sec:ch}

As demonstrated above, in a finite lattice general initial conditions evolve
into a chaotic state. For characterization of chaos we use Lyapunov
exponents. The chaotic state of the lattice has also been characterized in
  \cite{Sen-Mohan-Pfannes-04,Sen-Hong-Bang-Avalos-Doney-08} by the means of
  the velocity distribution of the lattice site. It has been found that the
  lattice possesses a quasi-non-equilibrium phase, characterized by a
  Boltzmann-like velocity distribution but without energy equipartition.

\begin{figure}
\centering
    \subfigure[]{
      \label{fig:lyap_b}
      \includegraphics[draft=false,angle=270,width=0.31\textwidth]{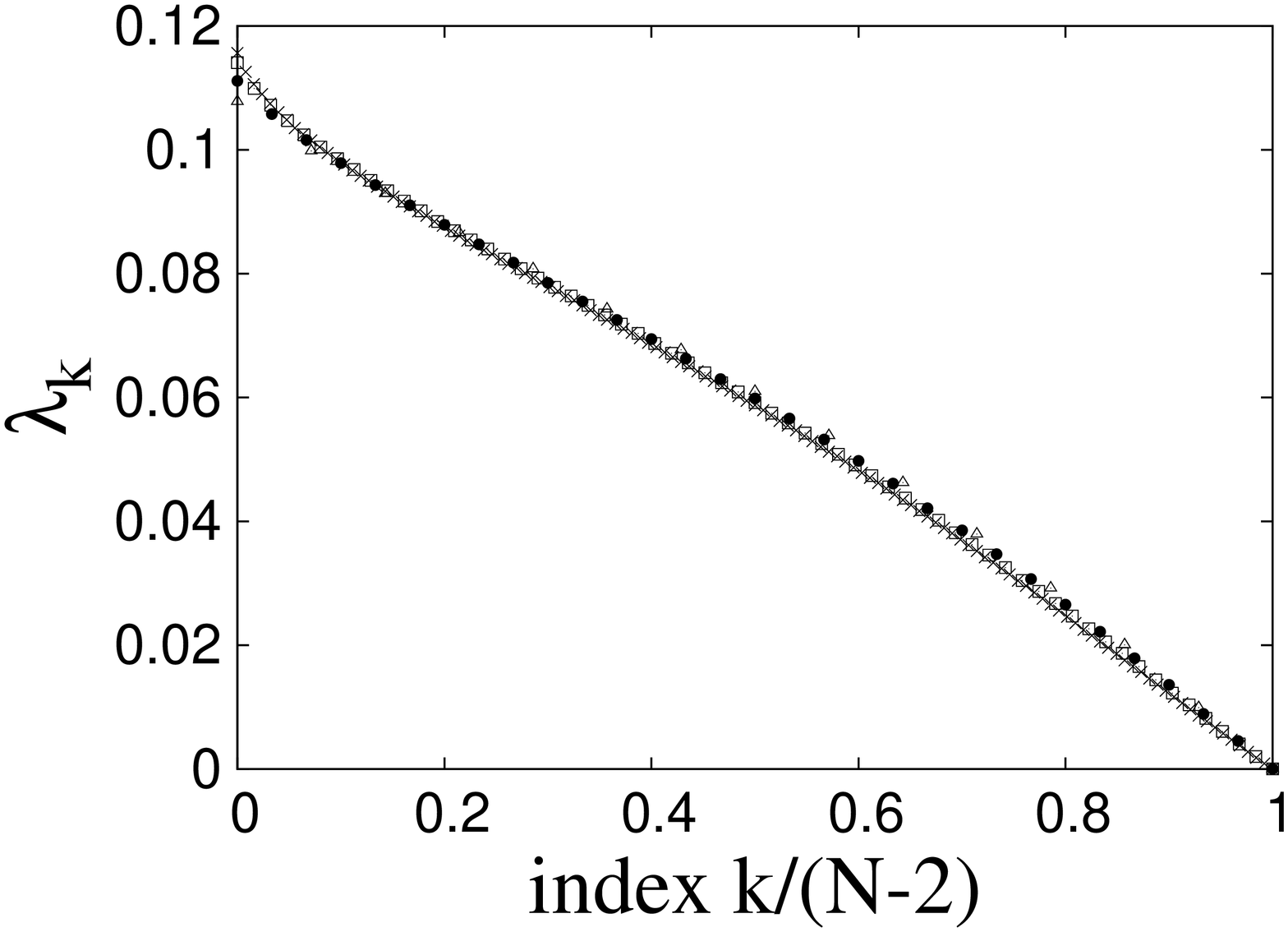}
    }
    \subfigure[]{
      \label{fig:lyap_a}
      \includegraphics[draft=false,angle=270,width=0.31\textwidth]{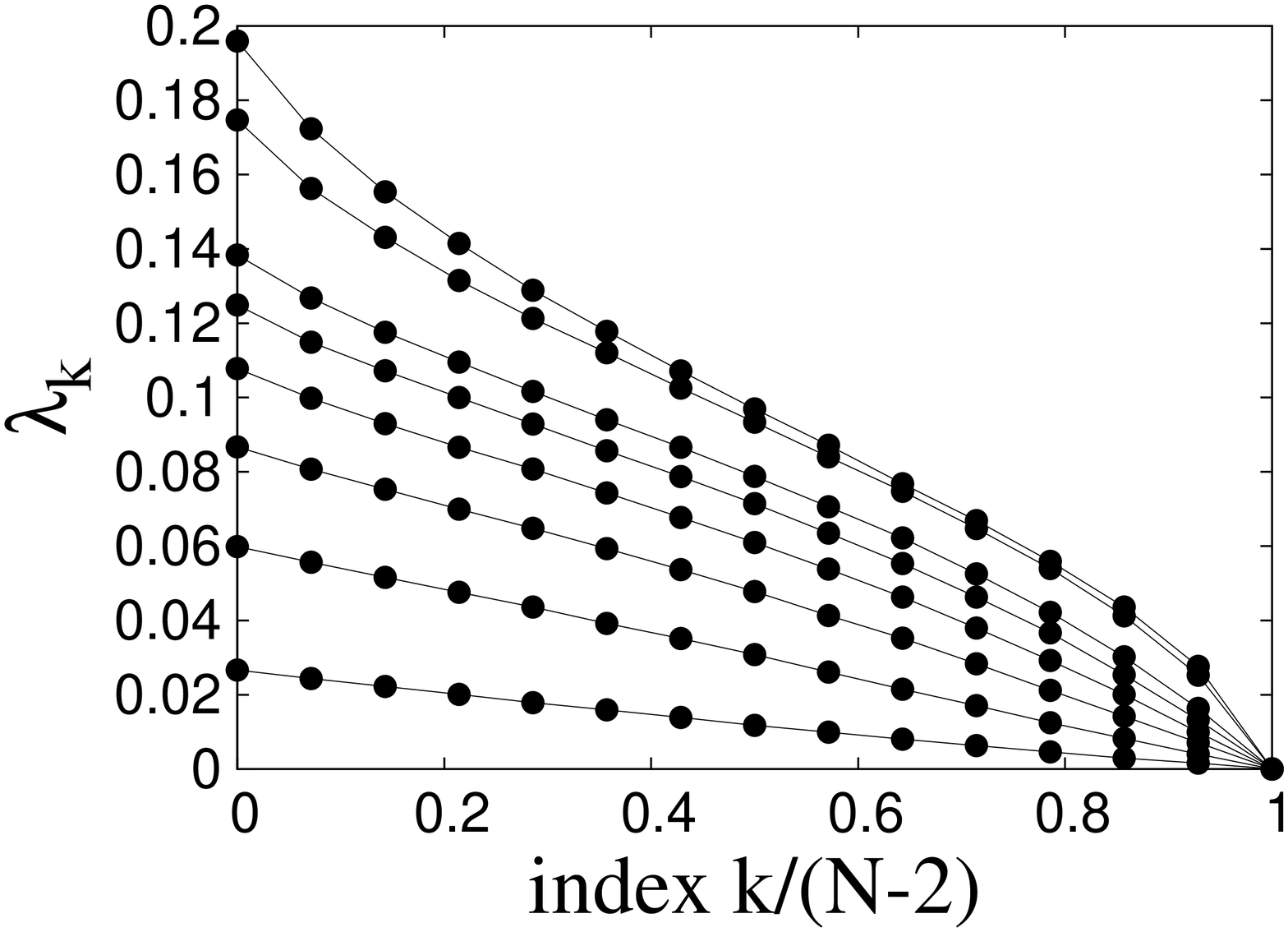}
    }
    \subfigure[]{
      \label{fig:lyap_c}
      \includegraphics[draft=false,angle=270,width=0.31\textwidth]{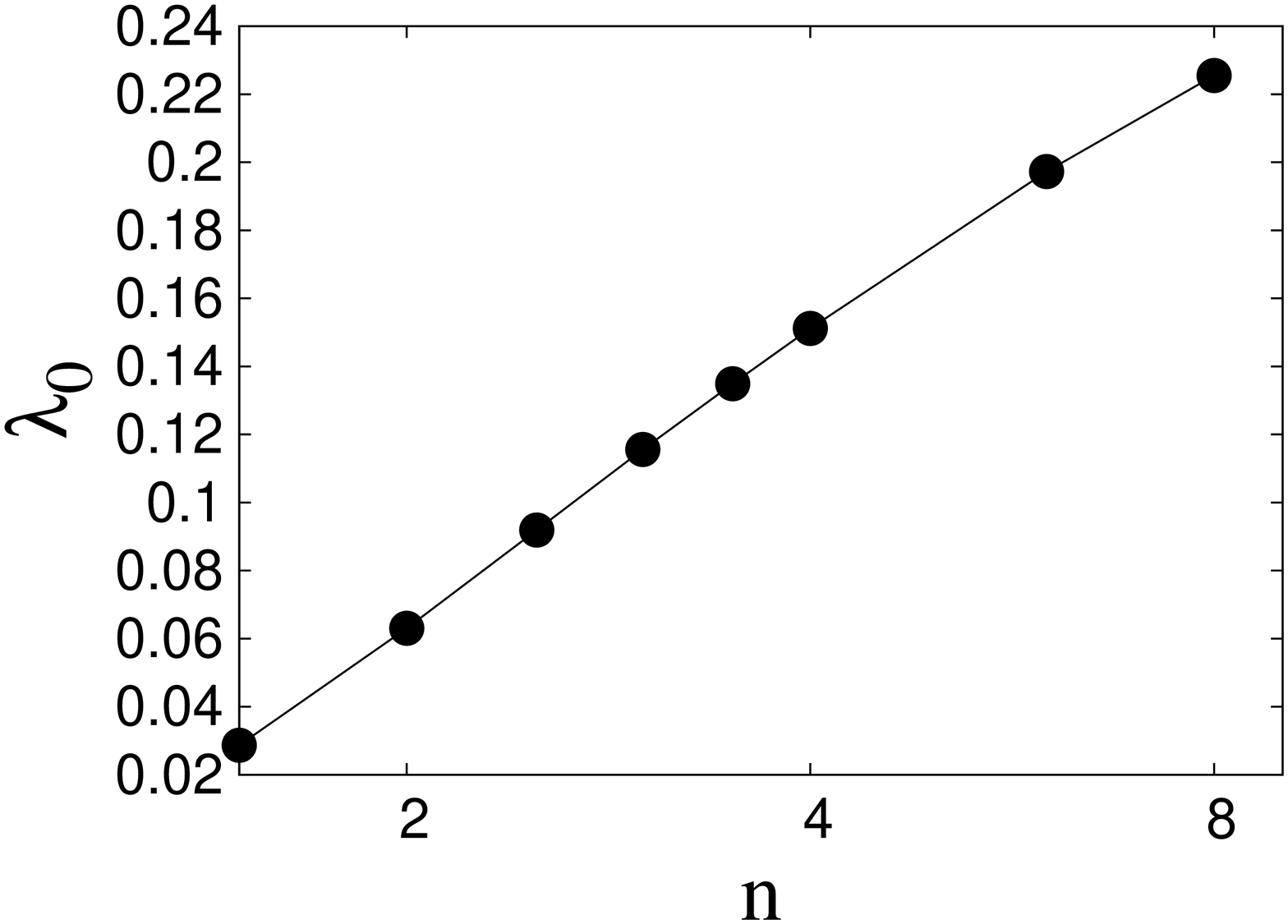}}
  \caption{Lyapunov exponents of the Hamiltonian \eref{eq:ham}. (a) The
    Lyapunov spectra for one fixed nonlinearity index $n=3$ and different
    values of lattice length $N=16,\,32,\,64,\,128$. The index axis is
    normalized to $1$. (b) The Lyapunov spectra $\lambda_j$ for various values
    of the nonlinearity index $n$ (from bottom to top:
    $n=1.5,\,2,\,2.5,\,3,\,3.5,\,4,\,6,\,8$) and fixed lattice length $N=16$.
    Larger values of $n$ produce stronger chaos that smaller ones.  (c) The
    largest Lyapunov exponent $\lambda_1$ for different values of $n$.  The
    horizontal axis is logarithmic, thus one can see that roughly $\lambda_0
    \sim \text{const}\cdot\log(n)$.}
  \label{fig:lyap}
\end{figure}

First, we check that chaos in the lattice is extensive, i.e. the
Lyapunov exponents form a spectrum when the system size becomes large
(Fig.~\ref{fig:lyap_b}). This property allows us to extend the calculations of
finite lattices to the thermodynamic limit. Note, that due to the two
conservation laws, four Lyapunov exponents vanish; we have not found any more
vanishing exponents, indicating the absence of further hidden conserved
quantities.

For a lattice of length $N=16$ the dependence of the Lyapunov exponents on the
nonlinearity is shown in Fig.~\ref{fig:lyap_a}. For a fixed total energy (we
have set $H=N=16$ in these calculations) the Lyapunov exponents grow with the
nonlinearity index. The plot presented in Fig.~\ref{fig:lyap_c} indicates that
$\lambda_{max}\propto \log n$, although we did not consider very high
nonlinearity indices to make a definite conclusion on the asymptotics for
large $n$.

We stress here that because of the scaling of the strongly nonlinear lattices
under consideration, chaos is observed for arbitrary small energies -- only
the Lyapunov exponents decrease accordingly.

\section{Conclusion}
\label{sec:con}

In this paper we have studied strongly nonlinear Hamiltonian lattices, with a focus on compact traveling waves and on
chaos. We have presented an accurate numerical scheme allowing one to find
solitary waves. Moreover, from the integral form representation used one
easily derives the super-exponential form of the tails.  In this way we have confirmed this remarkable result by Chatterjee~\cite{Chatterjee-99} by another analytical method and by accurate numerical analysis. The constructed
compactons were then studied via direct numerical simulations of the
lattice. Their collisions are nearly elastic, but the small non-elastic
components on a long time scale destroy the localized waves and result in a
chaotic state. Chaos appears to be a general statistically stationary state in
finite lattices, with a spectrum of Lyapunov exponents where the largest one
grows roughly proportional to the logarithm of the nonlinearity index.

We would like to mention here also several aspects that deserve further
investigations. Recently, a problem of heat transport in one-dimensional
lattices have attracted a large attention~\cite{Lepri-Livi-Politi-03}, here
the properties of strongly nonlinear lattices may differ from those possessing
linear waves. Also a quantization of these lattices seems to be a non-trivial
task, as there are no linear phonons to start with. Finally, the Anderson
localization property of disordered lattices has been recently intensively
discussed for nonlinear systems. For strongly nonlinear lattices the problem
has to be attacked separately, as here one cannot rely on the spectral
properties of a linear disordered system.

\acknowledgments

We thank P. Rosenau and D. Shepelyansky for constant stimulating
discussions. The work was supported by DFG via Grant PI-220/10 and via
Collaborative Research Project 555 ``Complex nonlinear processes''.


\end{document}